\documentclass[aps,pra,reprint,superscriptaddress]{revtex4-2}

\usepackage{graphicx}
\usepackage{color}

\usepackage{xspace}

\newcommand{\MBNExplorer}
{\textsc{MBN Explorer}\xspace}

\newcommand{\MBNStudio}
{\textsc{MBN Studio}\xspace}

\begin{document}

\title{Advances in multiscale modeling for novel and emerging technologies}

\author{Alexey V. Verkhovtsev}
\altaffiliation{On leave from Ioffe Institute, Polytekhnicheskaya 26, 194021 St. Petersburg, Russia}
\affiliation{MBN Research Center, Altenh\"oferallee 3, 60438 Frankfurt am Main, Germany}

\author{Ilia A. Solov'yov}
\altaffiliation{On leave from Ioffe Institute, Polytekhnicheskaya 26, 194021 St. Petersburg, Russia}
\affiliation{Department of Physics, Carl von Ossietzky Universit\"at Oldenburg, Carl-von-Ossietzky-Str. 9-11, 26129 Oldenburg, Germany}

\author{Andrey V. Solov'yov}
\altaffiliation{On leave from Ioffe Institute, Polytekhnicheskaya 26, 194021 St. Petersburg, Russia}
\affiliation{MBN Research Center, Altenh\"oferallee 3, 60438 Frankfurt am Main, Germany}

\begin{abstract}
Computational multiscale modeling encompasses a wide range of end-products and a great number of technological applications. This paper provides an overview of the computational multiscale modeling approach based on utilization of \MBNExplorer and \MBNStudio software packages, the universal and powerful tools for computational modeling in different areas of challenging research arising in connection with the development of novel and emerging technologies. Three illustrative case studies of multiscale modeling are reviewed in relation to (i) the development of novel sources of monochromatic high-energy radiation based on the crystalline undulators, (ii) controlled fabrication of nanostructures using the focused electron-beam induced deposition, and (iii) ion-beam cancer therapy. These examples illustrate the key algorithms and unique methodologies implemented in the software.
\end{abstract}

\maketitle


\section{Introduction}
\label{sec:Introduction}

Computational multiscale modeling encompasses a wide range of end-products and a significant number of applications in (i) the avionics and automobile industry for designing nanostructured materials, functionalized surface coating, as well as stronger and lighter materials for aircraft and cars; (ii) mechanical engineering for virtual design of superhard nanostructured materials;
(iii) medical applications for novel materials technologies for implants and tissue regeneration; (iv) electronic and chemical industry for constructing highly efficient batteries and catalyzers; (v) the pharmaceutical industry for drug design, etc.
In most of these applications, it is necessary to identify and design specific properties of the system determined by its molecular structure on the nanoscale and ensure the transfer of these properties to the macroscopic scale to make them functional and usable.
Such a transition implies multiscale modeling approaches that rely on the combined use of quantum mechanics
methods together with classical molecular dynamics (MD), or linkage of MD and Monte Carlo (MC) simulations,
or the application of efficient computational algorithms allowing to perform simulations across the scales.

This paper provides an overview of the computational multiscale modeling approach \cite{Solov'yov2017} based on utilization of the powerful and universal software packages \MBNExplorer \cite{Solovyov2012} and \MBNStudio \cite{Sushko2019}.
The paper begins with a brief overview of different theoretical methods for modeling Meso-Bio-Nano (MBN) systems and the limitations of these methods. It is followed by a brief description of the \MBNExplorer and \MBNStudio software packages and their application areas. Next, several case studies of multiscale modeling by means of \MBNExplorer and \MBNStudio are reviewed, illustrating the main algorithms and unique methodologies implemented in the software.

\begin{sloppypar}
There are many concrete examples of novel and emerging technologies benefiting from computational multiscale modeling.
As an illustration, a few of them are highlighted in the paper: (i) development of novel sources of monochromatic high-energy radiation based on the crystalline undulators, (ii) controlled fabrication of nanostructures using the focused electron-beam induced deposition (FEBID), and (iii) ion-beam cancer therapy.
These examples illustrate the new possibilities that computational multiscale modeling provides to novel and newly emerging technologies.
\end{sloppypar}

\section{Computational approaches in Meso-Bio-Nano Science}
\label{sec:approaches}

MBN Science is an interdisciplinary field of research studying structure formation and dynamics of animate and inanimate matter on the nano- and mesoscales. This field bundles up several traditional topics in theoretical physics and chemistry at the interface with life sciences and materials research under a common theme.

\begin{sloppypar}
The range of open challenging scientific problems in the field of MBN Science is very broad \cite{Solov'yov2017, ISACC_book_2008}. They may include: structure and dynamics of clusters, nanoparticles, biomolecules, and many other nanoscopic and mesoscopic systems; clustering, self-organization, growth and structure-formation processes, and their multiscale nature; assemblies of clusters/nanoparticles and bio-macromolecules; hybrid bio-nano systems; nanostructured materials; surface phenomena; nanoscale phase and morphological transitions; thermal, optical and magnetic properties; collective and many-body phenomena; electron transport and molecular electronics;
collision, fusion, fission and fragmentation processes; channeling effects; radiation effects; radiobiological effects, and many more.
\end{sloppypar}

\begin{figure}[t!]
\centering
\includegraphics[width=0.48\textwidth]{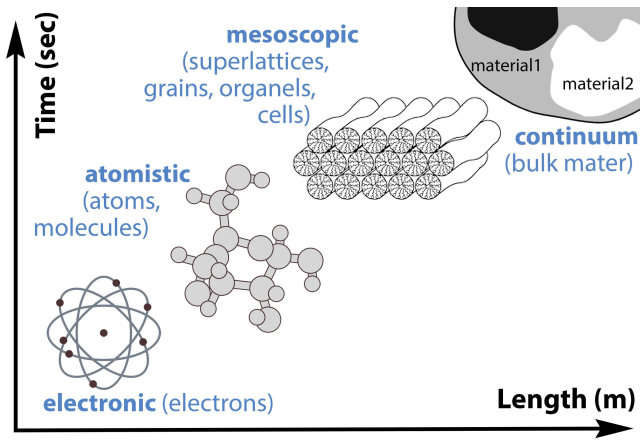}
\caption{An illustration of entities in the different computational techniques.}
\label{fig:multiscale_methods}
\end{figure}

MBN Science has been boosted immensely over the past two decades by the fast development of computer powers and related computational techniques that became broadly available.
This development resulted in a significant increase in the efficiency of available computer codes for scientific research.
Such codes are usually focused on particular systems, their particular sizes and phenomena involved (see Fig.~\ref{fig:multiscale_methods}),
and thus are limited in their ability to model physical, chemical or biological phenomena that go across the scales.
Therefore, much effort of research communities has been devoted to the computational approaches and modeling techniques to overcome this drawback and open new horizons in theoretical and computational research.

\subsection{Atomic and nano-scales}

\begin{sloppypar}
The characteristic quantum processes in MBN systems often involve electronic excitation, relaxation, fragmentation or transport dynamics.
Often, these effects in molecular and nanosystems occur when they are embedded into larger-scale molecular environments.
The quantitative description of the mentioned processes requires inclusion of both short-time/(sub-)nanoscale quantum
aspects and long-time/nano- and mesoscale environment effects.
Dynamical descriptions of molecular systems on the quantum atomic/subnano- and nano-/mesoscales are presently performed with disconnected theoretical and simulation tools.
There is no serious hope to explore in the near future both ranges within simply extending one tool to the other domain.
Bridging this gap implies interfacing quantum atomic/molecular and nano-/mesoscopic approaches.
These efforts will result from merging into a common model a time-dependent quantum mechanical (QM) approach and classical molecular dynamics (MD).
This will enable exploring new types of dynamics in a broad range of molecular systems, not accessible for analysis by other theoretical and computational means.
\end{sloppypar}

\begin{sloppypar}
Attempts have been made during recent years in various realizations of quantum mechanics/molecular mechanics (QM/MM) methods.
State-of-the-art achievements of these methods have been reviewed in several publications \cite{Senn_2009_AngChemIntEd.48.1198, Chung_2015_ChemRev.115.5678, Brunk_2015_ChemRev.115.6217}, and the field will experience a significant development in the future.
Many well-established quantum and classical MD codes,
being developed for a long time entirely independently, now reach the point
when they require coupling to go further across disciplines.
As this goal could not be achieved in a single universal approach, its realization often
requires the solution for each particular pair of quantum and classical codes with their
particular application areas and the communities of researches standing behind.
\end{sloppypar}

\subsection{Nano- and meso-scales}

\begin{sloppypar}
Description of structure and dynamics of molecular and nanosystems based on classical MD principles provides significant computational advantages compared to the quantum descriptions due to the simplicity of the formalism involved. Numerical realizations of this approach nowadays enable to treat the structures consisting of tens of millions of atoms and access dynamical time scales of molecular processes up to tens of microseconds, although for smaller systems, typically on the scale of hundreds of thousands of atoms.
Achievements made on the construction of software packages built on these principles have been reviewed in \cite{Sanbonmatsu2007, Zhao2013, Schulz_2009_JCTC.5.2798}.
The existing codes are based on the utilization of different classical force fields suitable for describing certain types of molecular systems, e.g. biomolecular systems, carbon or metallic nanosystems, etc.
However, many of the MD packages do not have the possibility (at least in the sufficiently elaborated form) to combine different force fields for the description of hybrid molecular systems and molecular environments due to the high level of optimization tuned for the specific type of force fields.
\end{sloppypar}

In recent years, it has become clear that a detailed understanding of numerous quantum molecular processes happening in larger-scale molecular environments becomes possible due to new advances in theoretical and experimental tools developed in atomic and molecular physics and nanotechnology.
An impressive example is the discovery of resonant mechanisms at the origin of radiation damage caused by low-energy electrons \cite{DNA2, DNA3} via the dissociative electron attachment (DEA) mechanism.
The identification of such a quantum effect in this context had an enormous impact: it paved the way to new research on possible low-energy radiation damage effects \cite{Surdutovich_2014_EPJD.68.353}.
Another example where the detailed knowledge of atomic and molecular processes in larger-scale molecular environments is crucial concerns novel radiosensitizers, e.g. metallic nanoparticles, which can enhance radiation effects on biological targets \cite{CNano_GNPs_review}.
An important feature of the aforementioned processes is that they occur on relatively short times (tens of femtoseconds) and in relatively small spatial domains (up to a few nanometers).

By exploiting the locality of the QM processes, one can isolate the active part of the system and describe it at the pure QM level.
However, in this case, possible large-scale effects coming from the other parts of the system might be missing.
For instance, this can be the case for a chromophore molecule in solvating shells or impurities in a solid.
The size of the QM system can be enlarged, but the convergence of such a strategy, in terms of accuracy and/or simulation time, can become slow.
Instead, one can explicitly include the environment and treat it classically at the MM level.
The development of such hierarchical techniques over past decades mainly operates in the field
of biological chemistry or surface science, where large sizes and/or environmental effects are crucial.
These methods allow one to tackle systems of thousands of atoms (a few tens at the QM level).

All-atom MD simulations involving very large systems may require significant computer resources that are not easily affordable. Similarly, simulations of processes on long timescales (beyond $\sim$1~ms) are prohibitively expensive because they require carrying out too many integration time steps. In these cases, one can sometimes tackle the problem by using reduced representations of the system where, instead of explicitly representing every atom of the system, groups of atoms are represented by ``pseudo-atoms''.
Such representations are sometimes called the coarse-grained models \cite{Kmiecik_2016_ChemRev.116.7898}. The parametrization of these models must be done empirically by matching the model's behavior to appropriate experimental data or all-atom simulations. Coarse-grained models have been used successfully to examine a wide range of questions in structural biology, liquid crystal organization, and polymer glasses \cite{Kmiecik_2016_ChemRev.116.7898}.

\subsection{Monte Carlo approach and finite element method}

Kinetic Monte Carlo (KMC) method is designed to model the time evolution of many-particle systems stepwise in time.
Instead of solving dynamical equations of motion, the KMC approach assumes that the system undergoes a structural transformation at each step of evolution with a certain probability.
The new configuration of the system is then used as the starting point for the next evolution step.
The transformation of the system is governed by several kinetic rates chosen according to the model considered.
Due to its probabilistic nature this methodology permits studying dynamical processes involving complex molecular systems on time scales significantly exceeding the characteristic time scales of conventional MD simulations \cite{Dick2011, 3DKMC, Moskovkin_2014_PSSB.251.1456}.
The KMC method is ideal in situations when certain minor details of dynamic processes become inessential,
and the major transition of the system to new states can be described by a few kinetic rates, determined
through the corresponding physical parameters.

\begin{sloppypar}
Quantitative structural analysis of macroscopically large systems and the description of some processes occurring on the macroscopic scale (such as heat transfer, fluid flow, mass transport, etc.) can be effectively carried out by means of finite element methods (FEM) \cite{Logan_FEM_book}. It is generally impossible to obtain analytical mathematical solutions for systems involving complicated geometries, loadings, and material properties.
FEM enables formulating the problem in terms of a set of algebraic equations rather than requiring the solution of differential equations. These numerical methods yield approximate values of the unknowns at discrete numbers of points in the continuum.
Hence this process of modeling a body by dividing it into an equivalent system of smaller bodies or units (finite elements) interconnected at points common to two or more elements (nodal points or nodes) and/or boundary lines and/or surfaces is called discretization.
In FEM, instead of solving the problem for the entire body in one operation, one formulates the equations for each finite element and combines them to obtain the solution of the whole body.
\end{sloppypar}

\section{\MBNExplorer and \MBNStudio}
\label{sec:MBN_E_S}

\begin{sloppypar}
Multiscale modeling of MBN systems is one of the hot topics of modern theoretical and computational research.
To fully understand and exploit all the richness and complexity of the MBN-world, especially its dynamics, one needs to consult many disciplines ranging from physics and chemistry to materials and life sciences, exploiting technologies from software engineering and high-performance computing. This general trend brought up the idea and then the development of \MBNExplorer \cite{Solovyov2012} and \MBNStudio \cite{Sushko2019}.
These software packages have been designed as powerful and universal instruments of computational research in the field of MBN Science, which are capable to explore, simulate, record and visualize both structure and dynamics of MBN systems, reproduce its known features and predict the new ones.
\end{sloppypar}

\textbf{MesoBioNano Explorer (\MBNExplorer)} \cite{Solovyov2012} is a software package for the advanced multiscale simulations of the structure and dynamics of various MBN systems. It has many unique features and a wide range of applications in physics, chemistry, biology, and materials science. It is suitable for classical non-relativistic and relativistic molecular dynamics (MD), Euler dynamics, reactive and irradiation-driven molecular dynamics (RMD and IDMD) simulations, as well as for stochastic dynamics or kinetic Monte Carlo (KMC) simulations of various randomly moving MBN systems or processes. These algorithms are applicable to a broad range of systems, such as nano- and biological systems, nanostructured materials, composite/hybrid materials, gases, plasmas, liquids, solids, and their interfaces, with sizes ranging from atomic to mesoscopic.

\subsection{Main features of \MBNExplorer}

\MBNExplorer enables calculations of energies of a large variety of MBN systems and optimization of their structures. The software package supports different types of molecular dynamics for MBN systems. The program operates with an extensive library of interatomic potentials, thus allowing to model many different molecular systems. It is also possible to simulate the dynamics of MBN systems in the presence of external fields -- electric, magnetic, gravitational, and electromagnetic waves.

Apart from the standard algorithms, \MBNExplorer is equipped with unique algorithmic implementations that significantly enhance the computational modeling capacities in various research and technological areas. The complete list of algorithms implemented in \MBNExplorer can be found in \cite{Solov'yov2017, Solo2017, Solov2017}.

In particular, \MBNExplorer has unique algorithmic implementations for multiscale modeling.
By means of stochastic dynamics (KMC, random walk dynamics) algorithms \MBNExplorer allows simulating the dynamics of MBN systems on time scales significantly exceeding the limits for the conventional atomistic MD simulations. Such a multiscale dynamics approach is ideal for systems in which details of their atomistic dynamics become excessive, and the overall behavior of a system can be reproduced through kinetic rates for the dominating modes of motion and probabilities of the key processes occurring in the system. This important feature of \MBNExplorer significantly expands its application areas and goes beyond the limits of other MD codes usually unable to deal with multiscale modeling.

\begin{sloppypar}
In the case of ultrarelativistic charged particle propagation through different media the implemented algorithms enable simulations of particle dynamics on macroscopically large distances with atomistic accuracy. These algorithms enable to obtain the necessary atomistic insights into macroscopically large systems and processes occurring therein. For instance, one can simulate the operation of novel intensive sources of high-energy monochromatic $\gamma$-rays based on irradiation of oriented crystals by beams of ultra-relativistic electrons and positrons.
\end{sloppypar}

\subsection{Main features of \MBNStudio}

In order to facilitate the practical work with \MBNExplorer a special multi-task software toolkit, called \textbf{\MBNStudio}, has been developed \cite{Sushko2019}. It simplifies modeling of MBN systems, setting up and starting calculations with \MBNExplorer, monitoring their progress and examining the calculation results. The software can be utilized for any type of calculations supported by \MBNExplorer.

\begin{sloppypar}
\MBNStudio enables the \textit{Project set-up} (standard as well as application-specific). Application-specific projects are usually designed for utilization in specific application areas, e.g. related to novel or emerging technologies. Such application-specific projects often involve special algorithms. A special modeling plug-in allows one to construct and prepare application-specific projects for simulation quickly and efficiently.
\end{sloppypar}

\MBNStudio has an advanced \textit{MBN system modeler}, a built-in tool for the computational design of MBN systems. By means of this plug-in one can easily assemble molecular systems of different geometries and compositions for their further simulations with \MBNExplorer.

\MBNStudio supports various standard \textit{input/output data formats} and links to numerous online \textit{databases and libraries} with coordinates and geometries for atomic clusters, nanoparticles, biomolecules, crystals and other molecular systems, which can be utilized in simulations with \MBNExplorer.

\MBNStudio is equipped with the \textit{output data handling, visualization and analytic tools} that allow calculation and analysis of specific characteristics determined by the output of MD simulations. Examples include calculations of diffusion coefficients, heat capacities, melting temperatures for solids, atomic radial distribution functions and many others.

\MBNStudio also enables \textit{video rendering} of the dynamics of MBN systems simulated with \MBNExplorer.

To summarize, \MBNExplorer and \MBNStudio are powerful tools for computational modeling in numerous challenging research areas arising in connection with the development of the aforementioned technologies. There are several such areas in which simulations performed by means of \MBNExplorer and \MBNStudio contributed immensely to their development. For instance, one of the areas concerns the development of novel light sources based on charged particles channeling in crystalline undulators \cite{Korol2020_LS_review, Channeling_book, Korol2021_EPJD_Collo}.
Another example concerns controllable fabrication of nanostructures with nanometer resolution using tightly-focused electron beams \cite{VanDorp2008, Utke_book_2012, Huth2012}. The third example deals with simulations of the nanoscopic molecular processes playing the key role in the ion-beam cancer therapy \cite{schardt2010heavy, Surdutovich_2014_EPJD.68.353, solov2016nanoscale}. \MBNExplorer combined with the visualization interface of \MBNStudio in many cases can substitute expensive laboratory experiments by computational modeling, making the software play a role of a ``computational nano- and microscope''. The following sections illustrate the capabilities of \MBNExplorer and \MBNStudio for advanced multiscale modeling in the three aforementioned technological areas.

\section{Computational modeling of novel crystal-based gamma-ray light sources}
\label{sec:CLS}

\begin{sloppypar}
\MBNExplorer and \MBNStudio provide unique possibilities for computational multiscale modeling of the physical processes and phenomena \cite{KSG1998, KSG_review_1999, KSG_review2004, MBNExplorer_Chan, Channeling_book, Korol2021_EPJD_Collo} that may facilitate the design and practical realization of novel gamma-ray Crystal-based Light Sources (CLS) \cite{Korol2020_LS_review}. Such light sources can be constructed through the exposure of oriented crystals (linear, bent, periodically bent) to beams of ultra-relativistic charged particles (e.g. electrons or positrons). The construction of novel CLSs is a challenging task that constitutes a highly interdisciplinary field entangling a broad range of correlated activities. CLSs provide a low-cost alternative to conventional X-ray LSs based on free electrons lasers (FEL) and have an enormous number of applications in the basic sciences, technology and medicine \cite{Korol2020_LS_review}.
\end{sloppypar}

The development of LSs for wavelengths $\lambda$ well below 1~\AA~(corresponding photon energies $E_{\rm {ph}} \gtrsim 10$~keV) is a challenging goal of modern physics. Sub-angstrom wavelength powerful spontaneous and, especially, coherent radiation will have many applications in nuclear and solid-state physics, and life sciences.
At present, several FEL facilities (FERMI, FLASH, LCLS, SACLA) provide femtosecond short laser-like photon pulses to user experiments. Their wavelengths range from the EUV and soft X-rays (FERMI, FLASH) to hard X-rays (LCLS, SACLA) \cite{LCLS2010, McNeilThompson_XFEL_2010}. However, no laser system has yet been commissioned for wavelengths $\lambda$ significantly shorter than 1~\AA~due to the limitations of permanent magnet and accelerator technologies.
Modern synchrotron facilities \cite{Ayvazyan_2002_EPJD, Yabashi2017_NatPhotonics} provide radiation of shorter wavelengths but of much less intensity which falls off very rapidly as $\lambda$ decreases.

Therefore, to create a powerful LS for $\lambda$ well below 1~\AA, i.e. in the hard X- and gamma-ray band, one should consider new approaches and technologies. Novel gamma-ray CLS can generate radiation in the photon energy range where the technologies based on the charged particles' motion in the fields of permanent magnets become inefficient or incapable. The limitations of conventional LSs are overcome by exploiting very strong crystalline fields that can be as high as $\sim$10$^{10}$~V/cm, which is equivalent to a magnetic field of 3000~T, whilst modern superconducting magnets provide $1-10$~T \cite{ParticleDataGroup2014}. The orientation of a crystal along the beam significantly enhances the strength of the particles' interaction with the crystal due to strongly correlated scattering from lattice atoms. This enables the guided motion of particles through crystals of different geometries and for the enhancement of radiation.

Practical realization of CLSs often relies on the channeling effect. The basic phenomenon of channeling is in a large propagation distance of a projectile particle that moves along a crystallographic plane or axis and experiences collective action of the electrostatic fields of the lattice atoms \cite{Lindhard}.
In the planar regime, positrons channel between two adjacent planes, whereas electrons propagate in the vicinity of a plane, thus experiencing more frequent collisions. As a result, a typical distance covered by a particle before it leaves the channeling mode due to uncorrelated collisions (the so-called dechanneling length) for positrons is much larger than for electrons.
To ensure enhancement of the emitted radiation due to the dechanneling effect, the crystal length must be chosen of the order of the dechanneling length \cite{KSG1998, KSG_review_1999, KSG_review2004}.

\begin{figure}[t!]
    \centering
    \includegraphics[width=0.48\textwidth]{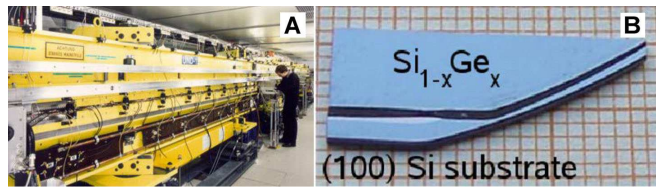}
    \caption{(\textbf{A}) Magnetic undulator for the European XFEL and (\textbf{B}) a Si$_{1-x}$Ge$_x$ superlattice Crystalline Undulator (CU) build atop the silicon substrate with the face normal to the $\langle 100 \rangle$ crystallographic direction \cite{Solov'yov2017}. The (110) planes  in the superlattice are bent periodically. The picture of a CU is courtesy of J.L. Hansen, A. Nylandsted and U. Uggerh{\o}j (University of Aarhus). }
    \label{fig:MagnUnd_CU}
\end{figure}

\begin{figure}[t!]
    \centering
    \includegraphics[width=0.48\textwidth]{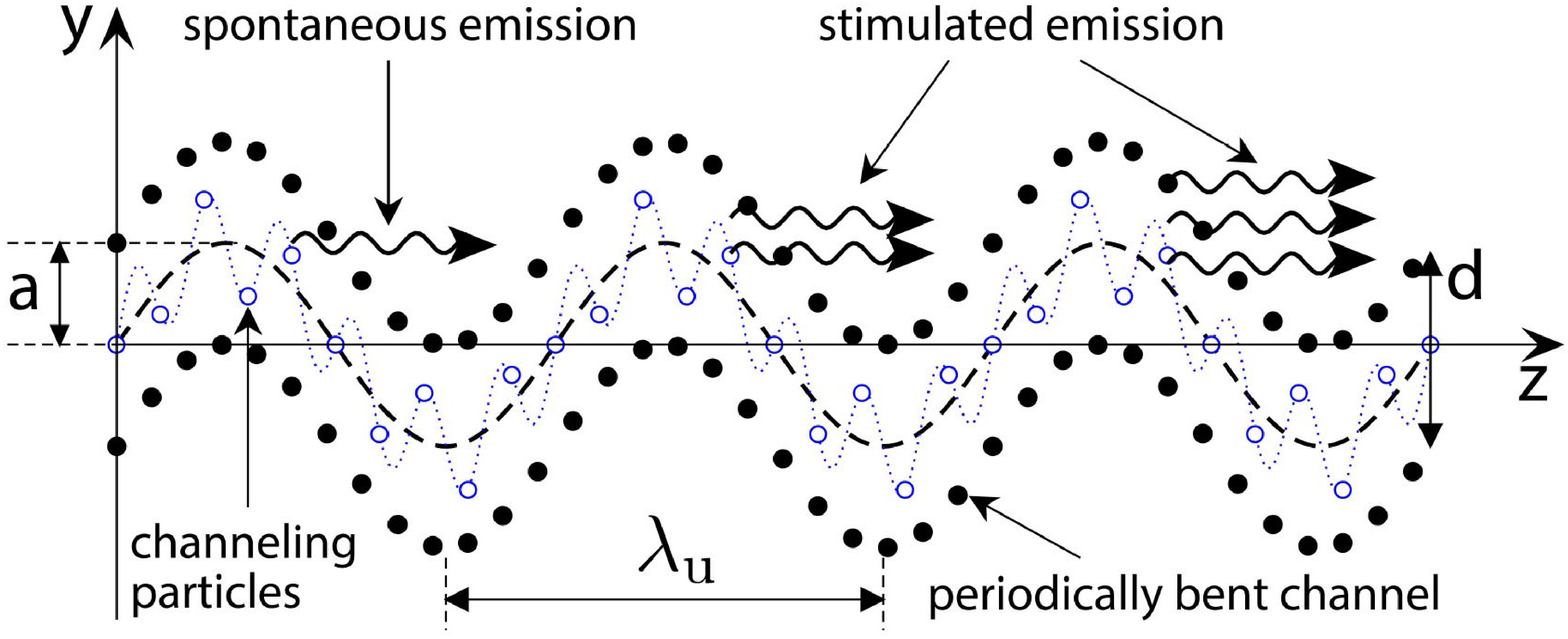}
    \caption{Schematic representation of a CU \cite{Channeling_book, KSG_review_1999, KSG_review2004}. Closed circles mark the atoms of crystallographic planes which are periodically bent with the amplitude $a$ and period $\lambda_{\textrm u}$. Thin dotted line illustrates the trajectory of the particle (open circles) which propagates along the center line (the undulator motion) and simultaneously undergoes so-called channeling oscillations. The periodic motion leads to the emission of the undulator-type radiation, and, under certain conditions, may result in the stimulated radiation.
    }
    \label{fig:CU_schematic}
\end{figure}

The projectile's motion and the radiation emission in bent and periodically bent crystals (BCs and PBCs) are similar to those in magnet-based synchrotrons and undulators. In the latter, the particles and photons move in vacuum. In contrast, in crystals, the particles propagate in a medium, leading to several limitations for the crystal length, bending curvature, and beam energy. However, the crystalline fields are so strong that they steer ultra-relativistic particles more effectively than the most advanced superconducting magnets. The strong crystalline fields enable to reduce the bending radius in BC down to the cm range and bending period $\lambda_{\textrm u}$ in PBCs (see Fig.~\ref{fig:CU_schematic}) to the hundred or even ten microns range.
The size and the corresponding cost of such devices are orders of magnitude smaller than that for the analogous devices based on magnets \cite{LCLS2010}.
Figure~\ref{fig:MagnUnd_CU} shows the comparison of the magnetic undulator for the European XFEL with the CU manufactured in the University of Aarhus and used in recent experiments \cite{Backe_EtAl_PRL_115_025504_2015}.

\begin{figure*}[t!]
    \centering
    \includegraphics[width=0.7\textwidth]{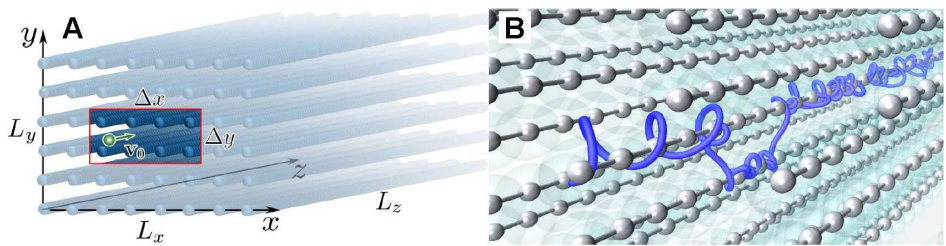}
    \caption{\textbf{A:} A schematic representation of the channeling process~\cite{MBNExplorer_Chan}. The $z$-axis is aligned with the incident beam direction (${\bf v}_0$ denotes the initial velocity) and is parallel to the crystallographic direction, along which the channeling is simulated. The $y$-axis is perpendicular to the crystal plane. At the entrance, the $x$ and $y$ coordinates of the particle are randomly chosen to lie in the central part of the ($xy$)-plane (the highlighted rectangle). \textbf{B:} Illustration of the trajectory of an axially channeling negatively-charged particle.
    }
    \label{fig:MBN_channeling}
\end{figure*}

One of the important practical realizations of the CLSs is based on the concept of a Crystalline Undulator (CU) \cite{Channeling_book, KSG1998, KSG_review_1999}. A CU device contains a PBC and a beam of ultra-relativistic positrons or electrons undergoing planar channeling, see Fig.~\ref{fig:CU_schematic}. In such a system, in addition to the channeling radiation (ChR) \cite{ChRad:Kumakhov1976}, the undulator radiation appears due to the periodic motion of the particles which follow the bending of the planes. A light source based on a CU can generate photons in the energy range from tens of keV up to the GeV region \cite{BezchastnovKorolSolovyov2014} (the corresponding wavelengths range starts at 0.1~\AA~and goes down to $10^{-6}$~\AA). The intensity and characteristic frequencies of the CU radiation (CUR) can be varied by changing the beam energy, the parameters of bending and the type of crystal. Under certain conditions a CU can become a source of the hard X- and gamma-ray laser light within the range $\lambda = 10^{-2} - 10^{-1}$~\AA~\cite{Channeling_book}, which cannot be reached in existing and planned FELs based on magnets.

The mechanism of the photon emission by means of CU is illustrated by Fig.~\ref{fig:CU_schematic} which presents a cross section of a single crystal. The $z$ axis is aligned with the mid-plane of two neighboring non-deformed crystallographic planes (not shown) spaced by the distance $d$. The closed circles denote the nuclei of the planes, which are periodically bent with the amplitude $a$ and period $\lambda_{\textrm u}$. The harmonic (sinusoidal) shape of periodic bending, $y(z) = a \sin{(2\pi z/\lambda_{\rm{u}})}$, is of a particular interest since it results in a specific pattern of the spectral-angular distribution of the radiation emitted by a beam of ultra-relativistic charged particles (the open circles in the figure) propagating in the crystal following the periodic bending.

The operational principle of a CU does not depend on the projectile type. Provided certain conditions are met, a particle undergoes channeling in a PBC \cite{Channeling_book, KSG1998, KSG_review_1999,  KSG_review2004}. Its trajectory contains two elements. First, there are channeling oscillations due to the action of the interplanar force. Their frequency $\Omega_{\rm{ch}}$ depends on the projectile energy $\varepsilon$, on the maximal value of the interplanar force and the interplanar distance $d$. Second, there are undulator oscillations due to periodicity of the bending whose frequency is $\Omega_{\rm{u}} \approx 2\pi c/\lambda_{\rm{u}}$. The spontaneous emission is associated with both of these oscillations. The typical frequency of ChR is $\omega_{\rm{ch}} \approx 2 \gamma^2 \Omega_{\rm{ch}}$, where $\gamma = (1 - v^2/c^2)^{-1/2}$ is the relativistic Lorenz factor with $v$ being the speed of a projectile and $c$ the speed of light. The frequency of CUR is $\omega_{\textrm u} \approx 2 \gamma^2 \Omega_{\textrm u}$. If $\Omega_{\textrm u} \ll \Omega_{\rm{ch}}$ then the ChR and CUR frequencies are well separated. In this case the characteristics of CUR are virtually independent on the channeling oscillations \cite{Channeling_book, KSG1998, KSG_review_1999}, and the operational principle of a CU is the same as of a magnet-based one (e.g. \cite{Ginzburg_1947, Motz_1951, RullhusenArtruDhez}) in which the monochromaticity of radiation is due to constructive interference of the photons emitted from similar parts of trajectory.

The feasibility for the construction of a PBC-based light source, a CU, was verified theoretically relatively recently
\cite{Dechan01, KSG1998, KSG_review_1999,  KSG_review2004, EnLoss00}. In the cited papers, as well as in the subsequent publications (see the latest review \cite{Channeling_book}), the essential conditions and limitations which must be met were formulated. These papers gave rise to coordinated theoretical, computational, technological and experimental studies of several related phenomena.

\begin{sloppypar}
\MBNExplorer enables simulations of propagation of various (positively and negatively charged, light and heavy) particles in various media, such as hetero-crystalline structures (including superlattices), bent and periodically bent crystals, amorphous solids, liquids, nanotubes, fullerites, biological environment, and many more. The applicability of the code to different structures can be adjusted either by choosing a proper interaction potential or, if necessary, by including a new potential \cite{MBNExplorer_Chan}.
\end{sloppypar}

By these means the channeling phenomenon \cite{Channeling_book, Lindhard}, which takes place when a charged particle enters a crystal at small angles with respect to a crystallographic direction, can be modeled \cite{Korol2021_EPJD_Collo}.
Depending on the orientation, one can distinguish planar and axial channeling when a projectile enters the crystal at small angles with respect to a crystallographic plane or axis, respectively.
The particle becomes confined and forced to move through the crystal, preferably along the crystallographic direction, experiencing collective action of the electrostatic field of the lattice ions (see Fig.~\ref{fig:MBN_channeling}). Since the field is repulsive for positively charged particles, they are steered into the interatomic region, while negatively charged projectiles move in close vicinity of ion strings or planes.
Figure~\ref{fig:MBN_channeling}B illustrates the axial channeling of a negatively charged projectile.

\subsection{Relativistic equations of motion}

In order to simulate motion of relativistic particles, \MBNExplorer considers \cite{MBNExplorer_Chan} relativistic equations of motion:
\begin{equation}
\left\{
\begin{array}{l l}
\textbf{\.{v}} = \displaystyle{ \frac{1}{m \gamma} \, \left( \textbf{F} - \textbf{v} \frac{\textbf{F} \cdot \textbf{v}}{c^2} \right) } & \ ,
\\
\displaystyle{ \textbf{\.{r}} = \textbf{v} } &
\end{array} \right.
\label{eq:rel_motion}
\end{equation}
where $\gamma$ is the relativistic Lorenz factor.
The force ${\bf F} = -{\bf \nabla} U({\bf r})$ acting on the projectile is due to the interaction with the crystal constituents. This interaction is characterized by the electrostatic potential energy $U({\bf r})$ which is considered as a sum of atomic potentials $U_{\rm at}$:
\begin{equation}
U({\bf r}) = \sum_j U_{\rm at}(|{\rm r} - {\rm R}_j|) \ ,
\end{equation}
where ${\rm R}_j$ is the position vector of the $j$th atom.
In \MBNExplorer the interaction between the charged projectiles and the constituent atoms is described by means of the widely used Moli\`{e}re approximation as well as the more recent approximation by Pacios \cite{Channeling_book}. Both interaction potentials are written as a product of the Coulomb potential and a screening function $\chi(r_{ij})$, see Refs.~\cite{Solov'yov2017, Channeling_book, Korol2021_EPJD_Collo} for details.

Applied to the propagation in a medium, equations~(\ref{eq:rel_motion}) describe the classical motion of a particle in the electrostatic field of the medium atoms.
As a first step in simulating the projectile's motion along a particular crystallographic direction, the crystalline structure inside the simulation box of the size $L_x \times L_y \times L_z$ is generated. The $z$-axis is aligned with the chosen crystallographic direction, see Fig.~\ref{fig:MBN_channeling}A. The integration of  equations of motion~(\ref{eq:rel_motion}) starts at $t=0$ when the particle enters the crystal at $z=0$. The initial coordinates $x_0$ and $y_0$ are chosen randomly within the central part of the ($xy$)-plane, shown by the red rectangle in Fig.~\ref{fig:MBN_channeling}A. The initial velocity ${\bf v}_0 = (v_{0_x}, v_{0_y}, v_{0_z})$ is predominantly oriented along the $z$-axis. The transverse components $v_{0_x}, v_{0_y}$ can be chosen with account for the beam emittance and energy distribution of particles in the beam \cite{MBNExplorer_Chan}.

\subsection{Dynamic simulation box}

\begin{sloppypar}
To simulate the propagation of particles through an infinite medium, \MBNExplorer utilizes the dynamic boundary conditions. In this case all particles in the system are separated into two groups: the moving particles and the fixed ones.
\end{sloppypar}

\begin{figure}[t!]
    \centering
    \includegraphics[width=0.48\textwidth]{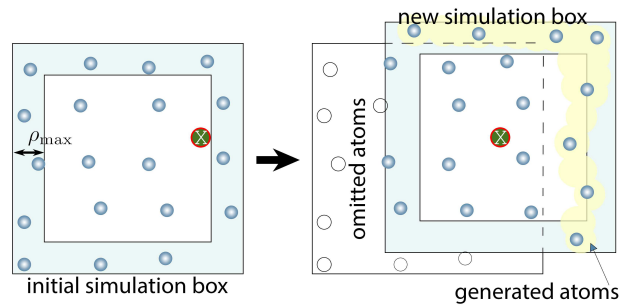}
    \caption{Illustration of the dynamic simulation box algorithm~\cite{MBNExplorer_Chan}. Once an X-marked projectile approaches the simulation box boundary (left panel), a new box of the same size is generated (right panel) with the particle placed approximately in its center. Positions of atoms (small shadowed circles) located in the intersection of the old and new boxes do not change. In the rest part of the new box the atomic positions are generated anew.
    }
    \label{fig:dynamic_simul_box}
\end{figure}

\begin{sloppypar}
For a single moving particle interacting with fixed atoms of the medium, the dynamic boundary algorithm is illustrated in Fig.~\ref{fig:dynamic_simul_box}. Once the particle approaches the border of the simulation box, a new box of the same size is generated with its center placed at the geometric center of the particle. To avoid spurious change in the force acting on the projectile, the positions of the atoms located in the intersection of the two boxes are conserved. The remaining part of the new box is filled with atoms of the medium following the specifications in the input file.
\end{sloppypar}

\begin{sloppypar}
The dynamic simulation box moves following the propagation of the particle. Thus, it allows the ``on-the-fly'' modeling of the crystalline structure in the course of integrating the relativistic equations of motion, Eq.~(\ref{eq:rel_motion}). Atoms of the medium are generated in the vicinity of the projectile as many times as it is necessary to propagate the projectile through the crystal of the thickness $L$. Periodic boundary conditions will not be effective in this case as the trajectory of the simulated particle is not periodic at all but instead rather unique at every instant. Similarly, one can consider the motion of many projectiles propagating through the medium built of fixed constituents. In this case, the computational scheme illustrated in Fig.~\ref{fig:dynamic_simul_box} is applied to each projectile.
\end{sloppypar}

\begin{figure*}[t!]
    \centering
    \includegraphics[width=0.8\textwidth]{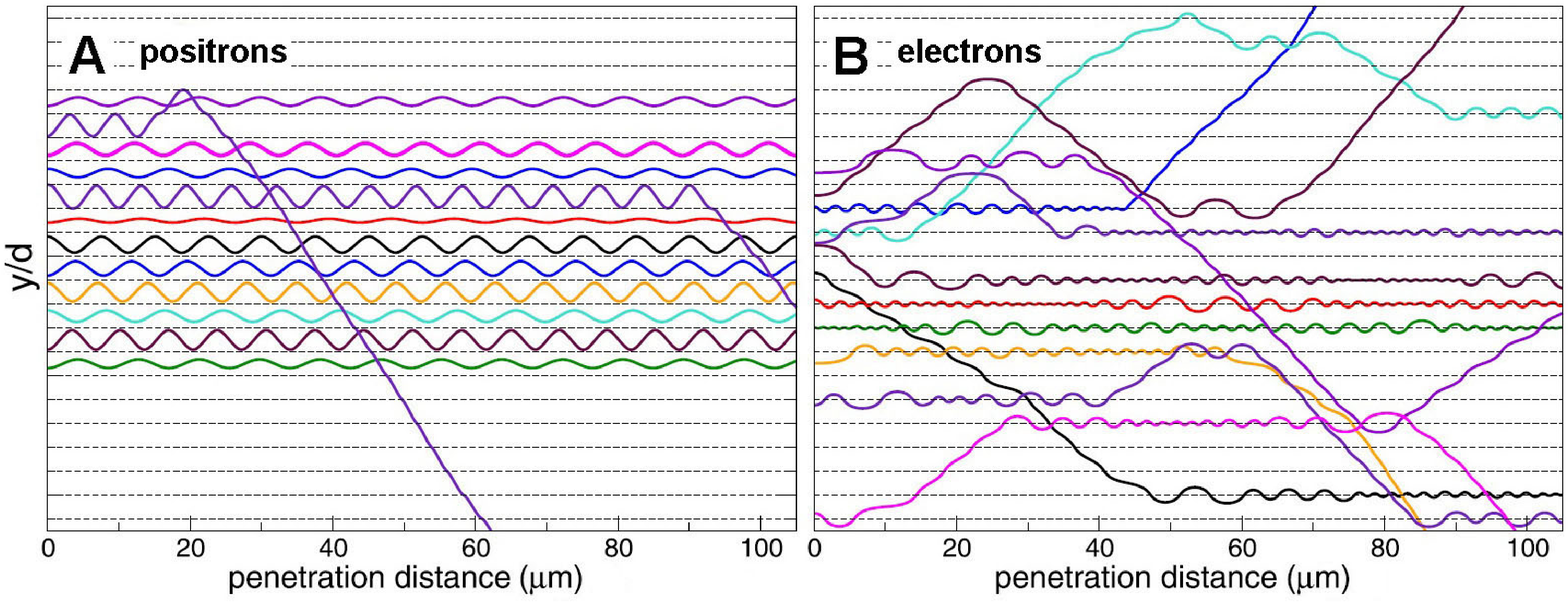}
    \caption{Channeling of 6.7 GeV positrons \textbf{(A)} and electrons \textbf{(B)} in a 105~$\mu$m thick silicon crystal. The plots show typical trajectories of the particles initially collimated along Si(110) crystallographic planes. The trajectories are simulated by means of the dedicated channeling module of \MBNExplorer \cite{MBNExplorer_Chan} employing the Moli\`{e}re potential. Horizontal dashed lines indicate the planes separated by the distance $d = 1.92$~\AA.}
    \label{fig:channeling_traj}
\end{figure*}

The dynamic simulation box algorithm permits to study at the atomistic level of detail the channeling phenomenon in mesoscopically large crystals, being micron-to-cm in length (see Fig.~\ref{fig:channeling_traj} as an illustrative example). Systems of such size cannot be handled by means of the all-atom MD approach with the standard periodic boundary conditions. Further details of this implementation are given in Refs.~\cite{MBNExplorer_Chan, Channeling_book}.

\MBNExplorer enables to simulate classical trajectories for different types of projectiles and their energies, different crystals (straight, bent and periodically bent) and crystallographic directions.
The latest version of the software enables accounting for the ionization energy losses and the radiation damping, as well as for the atom knockout processes.

\begin{sloppypar}
Figure~\ref{fig:channeling_traj} illustrates representative trajectories of 6.7~GeV positrons (panel~A) and electrons (panel~B) propagating in a 105~$\mu$m thick silicon crystal. The particles were initially collimated along Si(110) crystallographic planes. Horizontal dashed lines indicate the planes separated by the distance $d = 1.92$~\AA, that is the (110) interplanar distance in the silicon crystal at $T = 300$~K.
The trajectories shown in Fig.~\ref{fig:channeling_traj} have been selected from a bunch of simulated particle trajectories. In any particular case study, between $5 \times 10^3$ and $5 \times 10^4$ independent trajectories are typically considered. Each simulated trajectory describes the propagation of a single projectile through the medium. In the simulations, (i) transverse coordinates and velocities of the projectile at the crystal entrance and (ii) positions of the lattice atoms due to the thermal fluctuations are randomly sampled \cite{MBNExplorer_Chan}. Therefore, each simulated trajectory corresponds to a unique crystalline environment and all simulated trajectories are statistically independent.
\end{sloppypar}

Figure~\ref{fig:channeling_traj} illustrates a difference in channeling motion experienced by electrons and positrons. Positrons (Fig.~\ref{fig:channeling_traj}A) move along the (110) planes bouncing between two neighboring planes, and their oscillations exhibit nearly harmonic character. In contrast, electrons (Fig.~\ref{fig:channeling_traj}B) propagate oscillating in the vicinity of a plane. As a result, their trajectories are strongly non-harmonic and they tend to leave the channeling mode of motion earlier as compared to positrons.

Simulated trajectories can be used further for statistical characterization of the radiation emitted by a projectile. For each set of simulated trajectories of the total number $N_{\rm {traj}}$, \MBNExplorer provides an option to calculate the spectral distribution of the energy ${\rm d} E$ emitted within the cone $\theta \leq \theta_{0} \ll 1$ along the direction of the incident particle beam:
\begin{eqnarray}
\langle \frac{{\rm d} E}{{\rm d} (\hbar\omega)} \rangle
&=&
\frac{1}{N_{\rm {traj}}}
\sum_{n=1}^{N_{\rm {traj}}}
\frac{{\rm d} E_n }{ {\rm d} (\hbar\omega)} \ , \nonumber \\
\frac{{\rm d} E_n }{  {\rm d} (\hbar\omega)}
&=&
\int\limits_{0}^{2\pi}
{\rm d} \phi
\int\limits_{0}^{\theta_{0}}
\theta {\rm d}\theta\,
\frac{{\rm d}^2 E_n }{ {\rm d} (\hbar\omega)\, {\rm d}\Omega}.
\label{Methodology:eq.03}
\end{eqnarray}
Here, ${\rm d}^2 E_n / {\rm d} (\hbar\omega)\, {\rm d}\Omega$ stands for the spectral-angular distribution emitted by a projectile moving along the $j$th trajectory. In \MBNExplorer, this quantity is calculated within the framework of the quasi-classical approximation \cite{Baier} that combines the classical description of the particle's motion with the quantum corrections due to the radiative recoil and using the algorithm described in Refs.~\cite{MBNExplorer_Chan, Channeling_book}. The sum in Eq.~(\ref{Methodology:eq.03}) is carried out over all simulated trajectories, i.e. its takes into account the contribution of the channeling segments of the trajectories as well as of those corresponding to the non-channeling regime.

\begin{figure}[t!]
    \centering
    \includegraphics[width=0.42\textwidth]{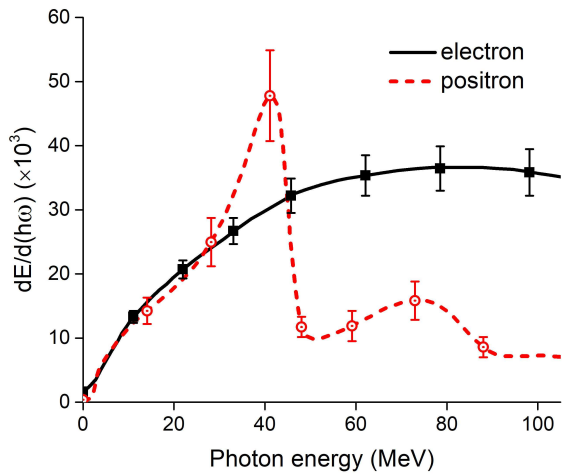}
    \caption{Radiation spectra from 6.7~GeV electrons (solid line) and positrons (dashed line) channeling through a 105~$\mu$m thick Si(110) crystal \cite{MBNExplorer_Chan}. The lines are drawn over ca. 200 photon energy points in which the spectra were calculated. The symbols mark a small fraction of the points and are drawn to illustrate typical statistical errors (due to a finite number of the simulated trajectories) in different parts of the spectrum.}
    \label{fig:rad_spectra}
\end{figure}

Figure~\ref{fig:rad_spectra} shows the dependence ${\rm d}E/{\rm d}(\hbar\omega)$ calculated for 6.7~GeV electrons and positrons aligned along Si(110) crystallographic plane at the crystal entrance \cite{MBNExplorer_Chan}. Statistical uncertainties due to the finite number of the analyzed trajectories ($\approx 500$ trajectories in each case) are indicated by the error bars which correspond to the probability $\alpha = 0.999$.

\begin{sloppypar}
The spectral distribution of emitted radiation, Eq.~(\ref{Methodology:eq.03}) is calculated by means of the quasi-classical formalism from the simulated trajectories. This means that all the radiation mechanisms associated with particle dynamics that are present in the system are accounted for.
The bremsstrahlung radiation is much weaker (by one-two orders of magnitude) than the channeling radiation providing particles move in the channeling regime \cite{Channeling_book}. The quasi-classical approximation \cite{Baier} used in the simulations reduces to the Bethe-Heitler theory (used to describe the bremsstrahlung in the energy range considered) if particles leave channels and experience incoherent interactions with atoms of the target crystal. The enhancement of the radiation in the channeling regime arises due to the coherent interaction of ultrarelativistic particles with the oriented crystalline planes or axes. These are all very well-known effects discussed e.g. in Refs.~\cite{Channeling_book, Korol2021_EPJD_Collo}. In the case of periodically bent crystals (as schematically shown in Fig.~\ref{fig:CU_schematic}), due to the periodicity of the particles' trajectories following the bending of the crystalline planes the photon emission spectrum attains the prominent features of the undulator radiation which are automatically captured by the quasi-classical formalism used in the simulations. These phenomena are described in detail in the book \cite{Channeling_book} and in the recent review papers \cite{Korol2020_LS_review, Korol2021_EPJD_Collo}.
\end{sloppypar}

In connection with the problems of constructing the high-quality undulator material and carrying out a quantitative analysis of the structures manufactured by different methods, \MBNExplorer enables all-atom MD simulations to model periodically bent structures under various external conditions. This approach, combined with modern numerical algorithms and advanced computational facilities, will bring the predictive power of the software up to the accuracy level comparable or higher than that achievable experimentally. It can turn computational modeling into a very useful tool, which in many cases could substitute expensive laboratory experiments by computational modeling and thus reduce the experimental and technological costs.

Verification of the channeling module of \MBNExplorer against experimental data as well as predictions of other theoretical models has been carried out in a large number of publications \cite{Korol2021_EPJD_Collo, MBNExplorer_Chan, BezchastnovKorolSolovyov2014, Sushko_AK_AS_2015, KorolBezchastnovSushkoSolovyov2016, Shen_2018_NIMB.424.26, Pavlov_2019_JPB.52.11LT01, Pavlov_2020_EPJD.74.21}, where channeling of electrons and positions in straight, bent and periodically bent crystals was studied, and the radiation spectra were calculated and successfully compared to experimental data.

\section{Computational modeling of the focused electron beam induced deposition process}

\begin{sloppypar}
The controllable fabrication of nanostructures with nanoscale resolution remains a considerable scientific and technological challenge \cite{Cui_Nanofabrication_book}. To address such a challenge, novel techniques have been developed \cite{Utke_book_2012} which exploit irradiation of nanosystems with collimated electron and ion beams. One of such techniques, Focused Electron Beam Induced Deposition (FEBID) \cite{VanDorp2008, Utke_book_2012, Huth2012}, is based on the irradiation of precursor molecules \cite{Barth2020_JMaterChemC} by high-energy electrons whilst they are being deposited on a substrate. Electron-induced decomposition releases the metallic component of the precursor which forms a deposit on the surface with a size similar to that of the incident electron beam (typically, a few nanometers) \cite{Utke2008}.
\end{sloppypar}

FEBID enables reliable direct-write fabrication of complex, free-standing 3D structures \cite{Huth2018, Utke2008}. Still, as the intended resolution falls below 10~nm, even FEBID struggles to yield the desired size, shape and chemical composition \cite{Utke2008, Thorman2015}, which primarily originates from the lack of molecular-level understanding of the irradiation-driven chemistry (IDC) underlying nanostructure formation and  growth \cite{Utke2008, Huth2012}. Computational multiscale modeling provides a new methodology for understanding IDC and, consequently, advancing controllable fabrication of nanostructures.

\begin{figure*}[t!]
    \centering
    \includegraphics[width=0.8\textwidth]{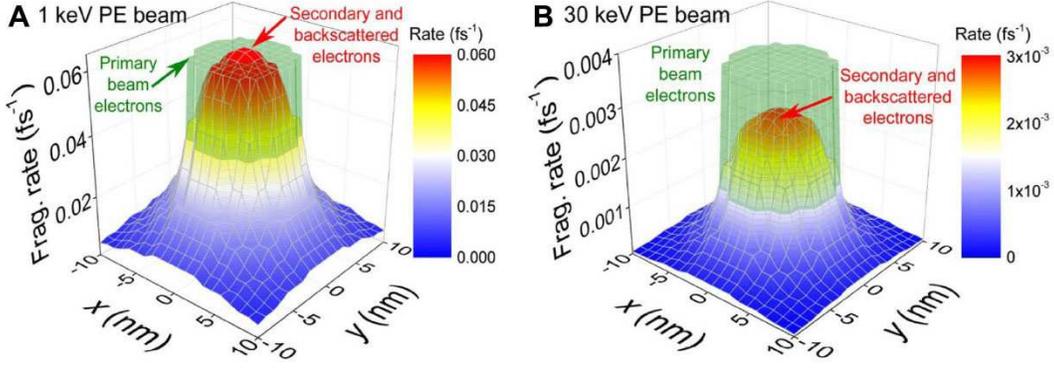}
    \caption{Electron-induced fragmentation rates for W(CO)$_6$ precursor molecules irradiated with PE beams of $E_0 = 1$~keV \textbf{(A)} and $E_0 = 30$~keV \textbf{(B)}. The green transparent surface depicts the PE beam area \cite{DeVera2020}.
    }
    \label{fig:FEBID_fragm_rate}
\end{figure*}

\begin{sloppypar}
FEBID operates through successive cycles of organometallic precursor molecules replenishment on a substrate and irradiation by a focused electron beam, which induces the release of metal-free ligands and the growth of metal-enriched nanodeposits. It involves a complex interplay of phenomena that require dedicated computational approaches:
(a) deposition, diffusion and desorption of precursor molecules on the substrate;
(b) multiple scattering of the primary electrons (PE) through the substrate, with a fraction of them being reflected (backscattered electrons, BSE) and the generation of additional secondary electrons (SE) by ionization;
(c) electron-induced dissociation of the deposited molecules; and
(d) the follow-up chemistry along with potential thermomechanical effects.
While processes (b) and (c) typically happen on the femtosecond-to-picosecond timescale, (a) and (d) may require up to microseconds or even longer.
\end{sloppypar}

Until recently, most computer simulations of FEBID and the nanostructure growth have been performed using a Monte Carlo (MC) approach and diffusion-reaction theory \cite{Utke_book_2012, Fowlkes2010, Sanz-Hernandez2017}, which allow simulations of the average characteristics of the process concerning local growth rates and the nanostructure composition. However, these approaches do not provide any molecular-level details regarding structure (crystalline, amorphous, mixed) and the IDC involved.
At the atomic/molecular level, ab initio methods permit the precise simulation of electronic transitions or chemical bond  reorganization \cite{Muthukumar2012, Muthukumar2018}, although their applicability is typically limited to the femtosecond–picosecond timescales and to relatively small molecular sizes. In between these approaches, classical MD \cite{Solov'yov2017} and particularly reactive MD \cite{Sushko_2016_EPJD.70.12} have proved to be very useful in the atomistic-scale analysis of molecular fragmentation and chemical reactions up to nanoseconds and  microseconds \cite{Sushko_2016_EPJD.70.12, deVera_2019_EPJD.73.215}.

\begin{sloppypar}
A breakthrough in the atomistic simulation of FEBID was achieved recently by means of Irradiation-Driven Molecular Dynamics (IDMD), a novel and general methodology for computer simulations of irradiation-driven transformations of complex molecular systems \cite{Sushko_IS_AS_FEBID_2016}. This approach overcomes the limitations of previously used computational methods and describes FEBID-based nanostructures at the atomistic level by accounting for chemical transformation of surface-adsorbed molecular systems under focused electron beam irradiation \cite{Sushko_IS_AS_FEBID_2016, Solov'yov2017, DeVera2020}.
\end{sloppypar}

Within the IDMD framework various quantum processes occurring in an irradiated system (e.g. ionization, bond dissociation via electron attachment, or charge transfer) are treated as random, fast and local transformations incorporated into the classical MD framework in a stochastic manner with the probabilities elaborated on the basis of quantum mechanics \cite{Sushko_IS_AS_FEBID_2016}.
Major transformations of irradiated molecular systems are simulated by means of MD with reactive CHARMM (rCHARMM) force field \cite{Sushko_2016_EPJD.70.12, Friis2020} using the \MBNExplorer \cite{Solovyov2012} and \MBNStudio \cite{Sushko2019} software packages.

\begin{sloppypar}
In the pioneering study \cite{Sushko_IS_AS_FEBID_2016} IDMD was successfully applied for the simulation of FEBID of W(CO)$_6$ precursors on a SiO$_2$ surface and enabled to predict the morphology, molecular composition and growth rate of tungsten-based nanostructures emerging on the surface during the FEBID process.
The follow-up study \cite{DeVera2020} introduced a novel multiscale computational methodology that couples event-by-event MC simulations of electron transport \cite{Dapor_2020_MC, Azzolini_2019_SEED} with IDMD for simulating the IDC processes with atomistic resolution.
The developed multiscale modeling approach enables simulation of radiation transport and effects in complex systems where all the FEBID-related processes (deposition, irradiation, replenishment) are considered. The spatial and energy distributions of secondary and backscattered electrons emitted from a SiO$_2$ substrate were used to simulate electron-induced formation and growth of metal nanostructures obtained after deposition of W(CO)$_6$ precursors on SiO$_2$.
\end{sloppypar}

\begin{sloppypar}
Within the IDMD framework, the space-dependent rate for bond cleavage in molecules on the substrate surface is given by:
\begin{eqnarray}
P(x,y) &=& \sigma_{\rm{frag}}(E_0) \, J_{\rm{PE}}(x,y,E_0) \nonumber \\
       &+& \sum_i \sigma_{\rm{frag}}(E_i) \, J_{\rm{SE/BSE}}(x,y,E_i) \ ,
\end{eqnarray}
where $E_0$ is the initial energy of the electron beam, $E_i < E_0$ a discrete set of values for the electron energies lower than $E_0$; $J_{\rm{ PE/SE/BSE}}(x, y, E_i)$ are space- and energy-dependent fluxes of PE/SE/BSE (electrons per unit area and unit time), and $\sigma_{\rm{frag}}(E_i)$ is the energy-dependent molecular fragmentation cross section. The PE beam flux at the irradiated circular spot of radius $R$ is:
\begin{equation}
J_0 = \frac{I_0}{e \, S_0} \ ,
\end{equation}
where $I_0$ corresponds to the PE beam current, $S_0 = \pi R^2$ to its area and $e$ is the elementary charge.
The electron distributions were simulated using the MC radiation transport code SEED (Secondary Electron Energy Deposition) \cite{Dapor_2020_MC, Azzolini_2019_SEED}. Molecular fragmentation and further chemical reactions were simulated by means of \MBNExplorer \cite{Solovyov2012} while \MBNStudio \cite{Sushko2019} was employed for constructing the molecular system, performing the precursor molecule replenishment phases, as well as for analyzing the IDMD simulation results.
\end{sloppypar}

\begin{figure}[t!]
    \centering
    \includegraphics[width=0.43\textwidth]{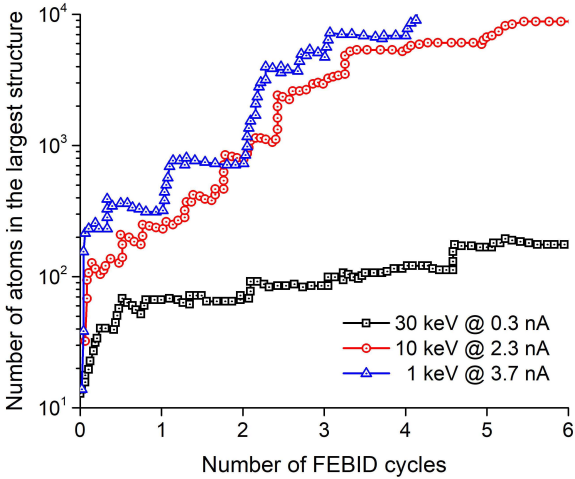}
    \caption{Evolution of the number of atoms in the largest simulated islands for PE beams of energies 1, 10 and 30~keV, for different currents as indicated \cite{DeVera2020}.
    }
    \label{fig:FEBID_cluster_size}
\end{figure}

Figure~\ref{fig:FEBID_fragm_rate} illustrates the space-dependent fragmentation rates induced by uniform 1 keV (panel~A) and 30 keV  (panel~B) beams of unit PE flux $J_0 = 1$ nm$^{-2}$fs$^{-1}$ within a circular area of radius $R = 5$~nm. Although the number of BSE/SE electrons for 30 keV is small, their large cross section (in relation to PE) produces a significant fragmentation probability, but less than that due to PE at the beam area. However, for 1 keV, the fragmentation probability due to BSE/SE ( $\sim 80-90$\% exclusively due to SE) is very large and significantly extends beyond the PE beam area. These results demonstrate the very different scenarios to be expected for beams of different energies, which will importantly influence the deposit properties and the prominent role of low-energy SE on molecular fragmentation.

Each irradiation phase lasts for a time known as dwell time, whose typical duration in experiments ($\ge \mu$s) is still computationally demanding for MD. To address this challenge, the irradiation phase was simulated for 10~ns and simulated PE fluxes $J_0$ (and hence PE beam currents $I_0$) were then scaled to match the same number of PE per unit area and per dwell time as in experiments \cite{Sushko_IS_AS_FEBID_2016}. As for replenishment, its characteristic times are also typically very long ($\sim$ms). In simulations, the CO molecules desorbed to the gas phase are removed during the replenishment stages and new W(CO)$_6$ molecules are deposited.

\begin{sloppypar}
As the irradiation-replenishment cycles proceed, atomic clusters and islands of different sizes and compositions appear on the substrate as the result of IDC, and the process of nucleation of metal-enriched islands and their coalescence starts \cite{Sushko_IS_AS_FEBID_2016}.
Figure~\ref{fig:FEBID_cluster_size} shows the number of atoms (either W, C or O) in the largest island for three simulation conditions close to reported in  experiments \cite{Porrati_2009_Nanotechnology}: 30~keV at $I_0 = 0.28$~nA, 10~keV at $I_0 = 2.3$~nA and 1~keV at $I_0 = 3.7$~nA. Smaller clusters tend to merge with time, giving rise to larger structures. The jumps in the island size observed with some frequency are due to the merging of independent clusters that grow on the substrate.
\end{sloppypar}

\begin{figure*}[t!]
    \centering
    \includegraphics[width=0.75\textwidth]{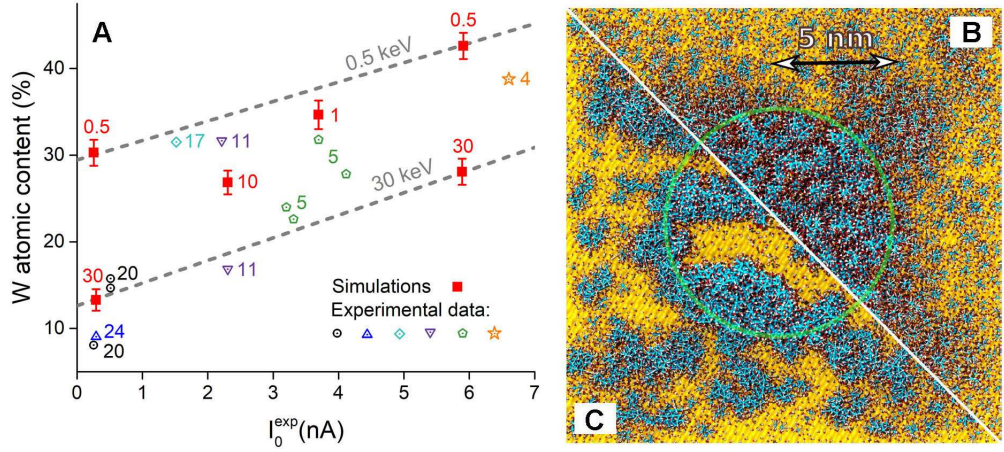}
    \caption{Compositions and morphologies of the deposits created by FEBID \cite{DeVera2020}. \textbf{(A)} Dependence of the deposit metal content on the beam energy $E_0$ and current $I_{\textrm{exp}}$ from experiments (open symbols) \cite{Porrati_2009_Nanotechnology} and simulations (full symbols) \cite{DeVera2020}. Numbers next to symbols represent the beam energy in keV for each case. Panels \textbf{B} and \textbf{C} show the top views of the deposits produced by 10keV@2.3nA and 1keV@3.7nA beams, respectively. The green area marks the PE beam spot while blue, white and red spheres represent, respectively, W, C and O atoms; the SiO$_2$ substrate is represented by a yellow surface.
    }
    \label{fig:FEBID_W-content}
\end{figure*}

Experimental measurements performed to date have been limited to particular values of energy and current due to the characteristics of the electron  source \cite{Porrati_2009_Nanotechnology}. In contrast, the IDMD simulation method permits the exploration of a much broader range of electron beam parameters.
Full symbols in Figure~\ref{fig:FEBID_W-content}A depict the simulated metal contents of the deposits as a function of experimentally equivalent current $I_{\rm{exp}}$. Error bars show the standard deviations obtained from three independent simulations for each case. Experimental  results \cite{Porrati_2009_Nanotechnology} are shown by open symbols. Numbers next to symbols represent the PE beam energies in keV. It is clearly seen that the results from simulations are within the range of experimental uncertainties, which indicates the predictive capabilities of the simulations.

This analysis provides a detailed ``map'' of the attainable metal content in the deposits as a function of the beam parameters, which is a valuable outcome for optimizing FEBID with W(CO)$_6$ on SiO$_2$. Dashed lines in Fig.~\ref{fig:FEBID_W-content}A correspond to the limiting values of PE beam energy and current studied. These results clearly show that, within the analyzed energy domain, a decrease in the beam energy and an increase in the current promote the faster growth of the deposit, as well as the augment in its metal content. Simulation results provide the grounds for clearly understanding such trends: an increment in the current means a larger number of PE per unit time, while a reduction in the energy produces an increase in the SE yield. These lead to both the greater size of the deposit and its larger metal content due to the increased probability for bond cleavage (see Fig.~\ref{fig:FEBID_fragm_rate}).

Figures~\ref{fig:FEBID_W-content}B and C show top views of the simulated deposits for 10keV@2.3nA and 1keV@3.7nA, after 7 and 5 irradiation cycles, respectively (the number of atoms in the largest island is approx. 12000 in both cases). The green circle marks the area covered by the PE beam (having a radius of 5 nm). These figures show that different energy-current regimes lead to distinct deposit microstructures and edge broadenings. While the more energetic 10~keV energy beam produces a deposit almost entirely localized within the PE beam area, the 1~keV beam produces a more sparse deposit (at least during the early stage of the FEBID process) that significantly extends beyond the PE beam area, producing an undesired edge broadening of the structure.

\begin{sloppypar}
The results presented in this Section demonstrate how the novel MC-IDMD approach provides the necessary molecular insights into the key processes behind FEBID, which can be used for its further optimization and development. The simulations \cite{Sushko_IS_AS_FEBID_2016, DeVera2020} have demonstrated a great predictive power, yielding fabricated nanostructure compositions and morphologies in good agreement with available experimental data \cite{Porrati_2009_Nanotechnology}.
The IDMD methodology provides a wide range of possibilities for atomistic-level study of FEBID and many other processes in which the irradiation of molecular systems and irradiation-driven chemistry play a key role.
\end{sloppypar}

\section{Computational modeling of ion-induced DNA damage in relation to ion-beam cancer therapy}

\begin{sloppypar}
\MBNExplorer and \MBNStudio can be utilized to evaluate radiobiological damage created by heavy ions propagating in different media, including biological. Such an analysis is behind the important biomedical technology known as the ion-beam cancer therapy (IBCT) \cite{schardt2010heavy, Linz2012_IonBeams, Surdutovich_2014_EPJD.68.353, solov2016nanoscale}.
IBCT allows delivery of high doses into tumors, maximizing cancer cell destruction and simultaneously minimizing the radiation damage of surrounding healthy tissue. The full potential of such therapy can only be realized if the fundamental mechanisms leading to lethal cell damage under ion irradiation are well understood. The key question is whether it is possible to quantitatively predict macroscopic biological effects caused by ion radiation on the basis of physical and chemical effects related to the ion-medium interactions on a nanometer scale.
\end{sloppypar}

\begin{figure}[t!]
    \centering
    \includegraphics[width=0.45\textwidth]{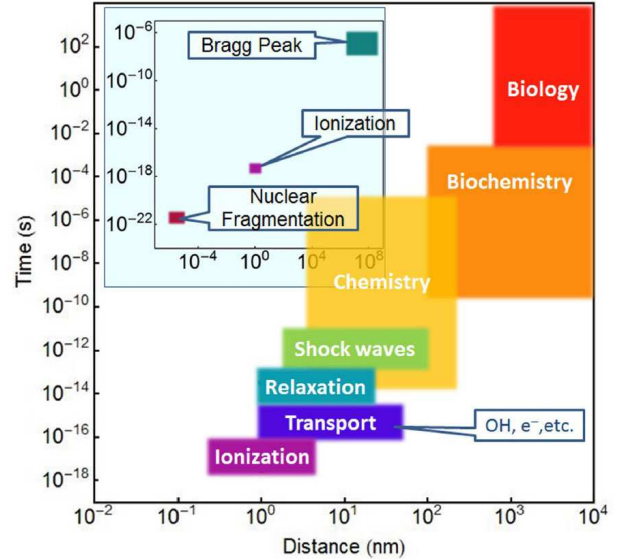}
    \caption{Features, processes, and disciplines, associated with radiation therapy, shown in a space-time diagram, which indicates temporal and spatial scales of the phenomena \cite{Surdutovich_2014_EPJD.68.353}. The history from ionization/excitation to biological effects are shown in the main figure, and features of ion propagation are shown in the inset. }
    \label{fig:MSA_scales}
\end{figure}

Recent review papers \cite{Surdutovich_2014_EPJD.68.353, surdutovich2019multiscale} and the book \cite{solov2016nanoscale} presented an overview of the main ideas of the MultiScale Approach to the physics of radiation damage with ions (MSA). This approach has the goal
of developing knowledge about biodamage at the nanoscale and molecular level and finding the relation between the characteristics of incident particles and the resultant biological damage.
The MSA is unique in distinguishing essential phenomena relevant to radiation biodamage at a given time, space and energy scale, and assessing the damage \cite{Surdutovich_2014_EPJD.68.353, surdutovich2019multiscale, solov2016nanoscale}. Temporal and spatial scales are schematically shown in Fig.~\ref{fig:MSA_scales}.

Radiation damage due to ionizing radiation is initiated by the ions incident on tissue. Initially, they have energy ranging from a few to hundreds of MeV per nucleon. In the process of propagation through tissue, the ions lose energy due to ionization, excitation, nuclear fragmentation, etc. Most of the energy loss of the ion is transferred to the tissue. Naturally, radiation damage is associated with this transferred energy, and the dose (i.e., deposited energy density) is a common indicator for the assessment of the damage \cite{Surdutovich_2014_EPJD.68.353, schardt2010heavy, Alpen}. The profile of the linear energy transfer (LET), that is the energy absorbed by the medium per unit length of the projectile's trajectory, along the ion's path is characterized by a plateau followed by a sharp Bragg peak.
The position of this peak depends on the initial energy of the ion and marks the location of the most severe damage in the target DNA molecule.
In the process of radiation therapy, a tumor is being ``scanned'' with the Bragg peak both laterally and longitudinally.

\begin{figure}[t!]
    \centering
    \includegraphics[width=0.45\textwidth]{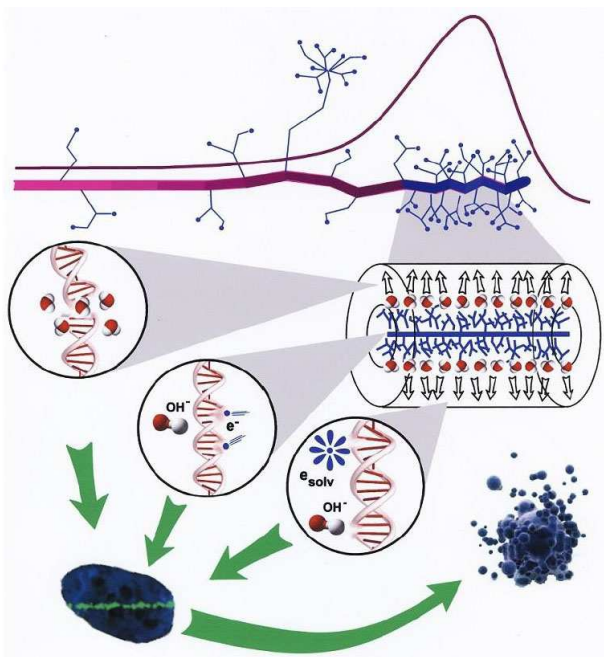}
    \caption{The scenario of biological damage with ions \cite{Surdutovich_2014_EPJD.68.353}. Ion propagation ends with a Bragg peak, shown in the top right corner. A segment of the track at the Bragg peak is shown in more detail. Secondary electrons and radicals propagate away from the ion's path damaging biomolecules (central circle). They transfer the energy to the medium resulting in the rapid temperature and pressure increase inside the cylinder. The ion-induced shock wave (shown in the expanding cylinder) due to the pressure increase damages biomolecules by stress (left circle), but it also effectively propagates reactive species to larger distances (right circle). A living cell responds to all shown DNA damage by creating foci (visible in the stained cells), in which enzymes attempt to repair the induced lesions. If these efforts are unsuccessful, the cell dies (the lower right corner). }
    \label{fig:MSA_scenario}
\end{figure}

However, the deposition of large doses in the vicinity of the Bragg peak does not explain how the radiation damage occurs, since projectiles themselves only interact with a few biomolecules along their trajectory and this direct damage is only a small fraction of the overall damage.
It is commonly understood that the secondary electrons and free radicals produced in the processes of ionization and excitation of the medium with ions are largely responsible for the vast portion of the biodamage.

\begin{figure*}[t!]
    \centering
    \includegraphics[width=0.70\textwidth]{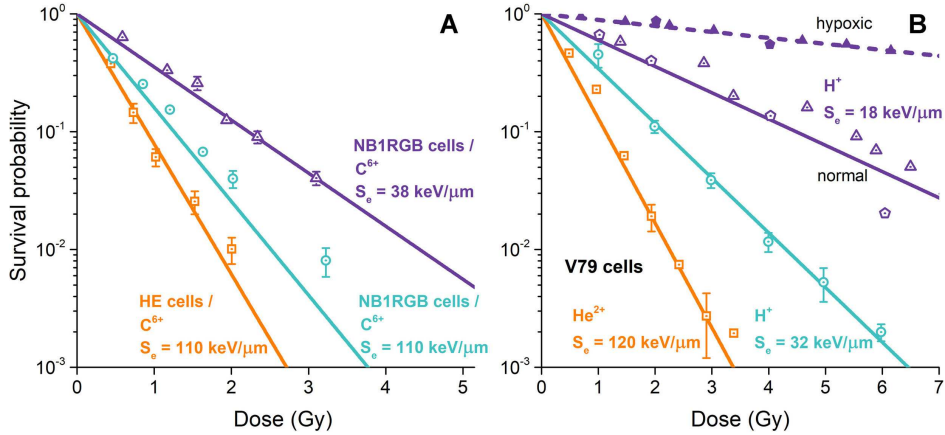}
    \caption{Survival probabilities as a function of deposited dose for different human \textbf{(A)} and Chinese hamster \textbf{(B)} cell lines. The survival probabilities calculated within the MSA \cite{Surdutovich_2014_EPJD.68.353} at the indicated values of LET are shown with lines. Experimental data for cells measured at a specific dose, either at normal or hypoxic conditions, are shown by symbols. See Refs.~\cite{verkhovtsev2016multiscale, verkhovtsev2019phenomenon} for further details.}
    \label{fig:MSA_cell_survival}
\end{figure*}

Secondary electrons are produced during a short time of $10^{-18} - 10^{-17}$~s following the ion's passage.
The energy spectrum of secondary electrons has been extensively discussed in the literature, see e.g. the review \cite{Surdutovich_2014_EPJD.68.353}. The main result is that most secondary electrons for ions at the Bragg peak region have energy below 40~eV. This has several important consequences.
First, the ranges of propagation of these electrons in tissue are rather small, below 10~nm~\cite{Meesungnoen02}.
Second, the angular distribution of their velocities, as they are ejected from their original host and as they scatter further,
is largely uniform \cite{Nikjoo06}; this allows one to consider their transport using a random walk approach \cite{precomplex, SurdutovichSolovyov_EPJD_2012, Solovyov2009_IBCT, epjdmarion}.

The next time scale $10^{-16} - 10^{-15}$~s corresponds to the propagation of secondary electrons in tissue.
In liquid water, the mean free paths of elastically scattered and ionizing 50-eV electrons are about 0.43 and 3.5~nm, respectively \cite{Nikjoo06}. This means that they ionize a molecule after about seven elastic collisions, while
the probability of second ionization is small \cite{epjd}. Thus, the secondary electrons are losing most of their energy within first 20 collisions and this happens within $1-2$~nm of the ion's path \cite{surdutovich2013biodamage}.
After that they continue propagating, elastically scattering with the molecules of the medium until they get bound or solvated electrons are formed. It is important to notice that these low energy electrons remain important agents for biodamage since they can attach to biomolecules like DNA causing dissociation \cite{Sanche11}. The solvated electrons may play an important role in the damage scenario as well \cite{hyd2, Kohanoff2012}.

The energy lost by secondary electrons in the processes of ionization and excitation of the medium is transferred to its heating (i.e. vibrational excitation of molecules) due to the electron-phonon interaction.
As a result, the medium within a $1 - 2$-nm region (for ions not heavier than iron) surrounding the ion's path is heated up rapidly \cite{Toulemonde2009_PRE, surdutovich2013biodamage}.
The pressure inside this narrow region increases by several orders of magnitude (e.g. by a factor of 10$^3$ for a carbon ion at the Bragg peak \cite{Toulemonde2009_PRE}) compared to the pressure in the medium outside that region.
This pressure builds up by about $10^{-14}-10^{-13}$~s and it is a source of a cylindrical shock wave \cite{surdutovich2010shock} which propagates through the medium for about $10^{-13} - 10^{-11}$~s. Its relevance to the biodamage is as follows.
If the shock wave is strong enough (the strength depends on the distance from the ion's path and the LET), it may inflict damage directly by breaking covalent bonds in a DNA molecule \cite{Yakubovich_2011_AIP.1344.230, surdutovich2013biodamage, devera2016molecular, fraile2019first, bottlander2015effect, Friis2020, Friis2021_SWdamage}.
Besides, the radial collective motion of the medium induced by the shock wave is instrumental in propagating the highly reactive molecular species, such as hydroxyl radicals and solvated electrons, to large radial distances (up to tens of nanometers) thus increasing the area of an ion's impact \cite{Friis2021_SWdamage, Surdutovich_2015_EPJD.69.193, deVera_2018_EPJD.72.147}.

\begin{sloppypar}
The assessment of the primary damage to DNA molecules and other parts of cells due to the above effects is done within the MSA.
This damage consists of various lesions on DNA and other biomolecules. Some of these lesions may be repaired by the living system, but some may not and the latter may lead to cell inactivation.
\end{sloppypar}

The scenario described above is illustrated in Fig.~\ref{fig:MSA_scenario} \cite{Surdutovich_2014_EPJD.68.353}.
The detailed comparison of its outcomes with experimental observations on the survival probability of irradiated cells was performed in Refs.~\cite{verkhovtsev2016multiscale, verkhovtsev2019phenomenon}. Figure~\ref{fig:MSA_cell_survival} highlights this comparison for several exemplar case studies.

\subsection{Simulation of the thermomechanical damage of the DNA by the ion-induced shock wave}

The direct thermomechanical damage of the DNA molecule as a result of interaction with the ion-induced shock wave has been explored using \MBNExplorer \cite{Yakubovich_2011_AIP.1344.230, surdutovich2013biodamage, devera2016molecular, Friis2020, Friis2021_SWdamage}.
In the earlier investigations \cite{Yakubovich_2011_AIP.1344.230, surdutovich2013biodamage, devera2016molecular}, the DNA damage by ion-induced shock waves was studied by means of classical MD simulations using non-reactive molecular mechanics force fields.
In those simulations the potential energy stored in a particular DNA bond was monitored in time as the bond length varied around its equilibrium distance \cite{surdutovich2013biodamage, devera2016molecular}.
When the potential energy of the bond exceeded a given threshold value, the bond was considered broken.

\begin{figure*}[t!]
    \centering
    \includegraphics[width=0.75\textwidth]{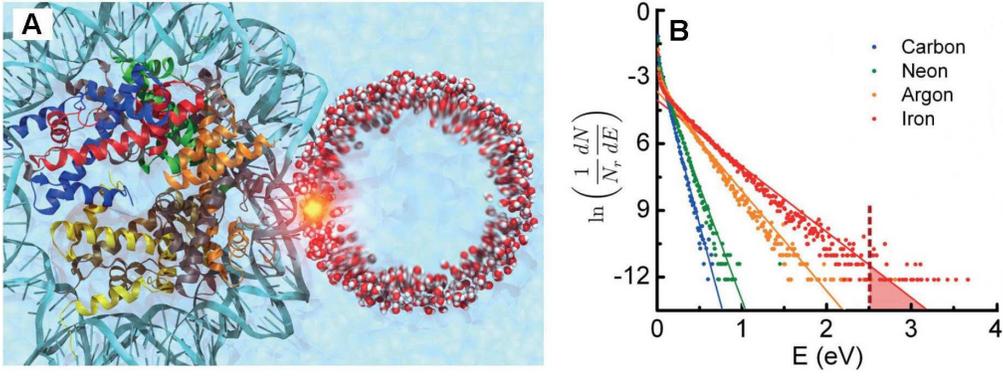}
    \caption{\textbf{A:} The cylindrical shock wave front in water (on the right; ion's path is the axis of this cylinder, perpendicular to the figure plane) interacts with a segment of a DNA molecule on the surface of a nucleosome (on the left). The yellow dot indicates the place where damage occurs. The medium is very dense following the wave front and is rarefied in the wake.
    \textbf{B:} The dependence of the logarithm of the normalized number of the covalent bond energy records for the selected DNA backbone region per 0.01 eV energy interval on the bond energy for four values of LET: 900, 1730, 4745, and 7195~eV/nm, corresponding to the Bragg peak values for carbon, neon, argon, and iron ions, respectively. Straight lines correspond to the fits of these distributions. The figures are adapted from Ref.~\cite{surdutovich2013biodamage}. }
    \label{fig:MSA_SW_nucleosome}
\end{figure*}

\begin{sloppypar}
In the pioneering study \cite{surdutovich2013biodamage} the MD simulations were focused on the interaction of the cylindrical shock wave originating from ion's path with a fragment of a DNA molecule situated on the surface of a nucleosome, see Fig.~\ref{fig:MSA_SW_nucleosome}A. Nucleosomes, histone-protein octamers wrapped about with a DNA double helix, are the primary structural units of chromatin, which is a principal component of the cell nucleus in eukaryotic cells. The simulations \cite{surdutovich2013biodamage} were done for four values of LET, namely 900, 1730, 4745 and 7195 eV/nm, corresponding to the Bragg peak values for carbon, neon, argon and iron ions, respectively.
Carbon ions are clinically used for cancer treatment, whereas heavier ions up to iron are present in galactic cosmic rays, being potentially damaging for humans during space missions \cite{durante2011physical, Kronenberg2012_HealthPhys.103.556}.
\end{sloppypar}

\begin{sloppypar}
The simulations were performed using the standard CHARMM force field \cite{CHARMM} which implies the harmonic approximation for describing the interaction potentials for covalent bonds and thus does not allow to observe bond breaking events directly. Therefore, to study whether the covalent bonds in the DNA backbone can be broken during the shock wave action, the energy temporarily deposited to these bonds was calculated. The analysis of MD simulations \cite{surdutovich2013biodamage} performed for the four indicated values of LET gives the distributions of the bond energy records. These records can be represented by a histogram that assigns to every interval of energy ($\varepsilon, \varepsilon + \delta \varepsilon$) the number of records corresponding to the bond energies from this interval. For each value of LET, the bond energy distribution was constructed. These distributions (normalized to the total number of records $N_r$ for each value of LET) are shown in Fig.~\ref{fig:MSA_SW_nucleosome}B, where $\ln{(1/N_r dN/d\varepsilon)}$ is plotted versus the corresponding energy interval.
Next, the number of energy records of selected covalent bonds of the DNA backbone exceeding a given threshold was counted. This was done by direct counting of bond energy records for which $E > E_0$, where $E_0$ is a variable threshold. The records counted for $E_0 = 2.5$~eV are shown in Fig.~\ref{fig:MSA_SW_nucleosome}B to the right of the dashed vertical line.
\end{sloppypar}

\begin{sloppypar}
A more quantitative description of the ion-induced shock wave phenomenon has become possible \cite{Friis2021_SWdamage} by means of reactive MD simulations that permit explicit simulation of covalent bond rupture and formation \cite{Sushko_2016_EPJD.70.12}. A recent study \cite{Friis2020} presented a detailed computational protocol for modeling the shock wave induced DNA damage by means of the rCHARMM force field \cite{Sushko_2016_EPJD.70.12}.
\end{sloppypar}

The target DNA molecule studied in \cite{Friis2020, Friis2021_SWdamage} contained 30 complementary DNA base pairs. The molecule was placed in a water box extending 17~nm from the DNA in the $x$- and $y$-directions and 8~nm in the $z$-direction. The total system size was 1,010,994 atoms.

\begin{figure*}[t!]
    \centering
    \includegraphics[width=0.67\textwidth]{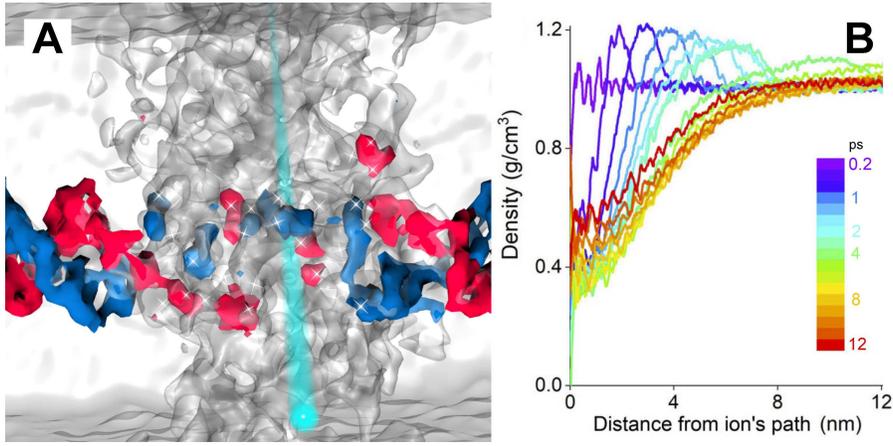}
    \caption{\textbf{A:} Illustration of the propagation of the ion-induced shock wave in the vicinity of the DNA segment containing 30 base pairs \cite{Friis2020}. Water molecules inside the 1-nm radius ``hot'' cylinder surrounding the ion track (shown by the yellow arrow) are highlighted. \textbf{B:} The density of water in the radial direction from the ion's path is shown at different instants from 0 to 12~ps after irradiation. }
    \label{fig:MSA_SW_DNA-density}
\end{figure*}

\begin{sloppypar}
It is widely established that one of the key events of radiation-induced DNA damage concerns the formation of single- and double-strand breaks (SSBs and DSBs) of the sugar-phosphate backbone.
Therefore, the rCHARMM force field was used to describe interatomic interactions in the C$_3^{\prime}$--O, C$_4^{\prime}$--C$_5^{\prime}$, C$_5^{\prime}$--O and P--O bonds in the DNA backbone, which connect the sugar ring of one nucleotide and the phosphate group of an adjacent nucleotide.
Bond dissociation energies and cutoff distances for bond breakage/formation were determined from density functional theory (DFT) calculations \cite{Friis2020}. Covalent interactions in other parts of the DNA molecule were modeled using the standard CHARMM force field.
\end{sloppypar}

As described above, low-energy electrons with the average kinetic energy of about 40~eV, which are predominantly produced for ions at the Bragg peak region, lose most of their energy by ionizing and exciting molecules of the medium within approximately 1 nanometer from the ion's path \cite{Surdutovich_2014_EPJD.68.353, Surdutovich_2015_EPJD.69.193}. Therefore, in the MD simulations the energy lost by the propagating ion has been deposited into the kinetic energy of water molecules located inside a ``hot'' cylinder of 1~nm radius around the ion's path.
The equilibrium velocities of all atoms inside that spatial region are increased by a factor $\alpha$ such that the kinetic energy of these atoms reads as \cite{Yakubovich_2011_AIP.1344.230, Yakubovich_2012_NIMB.279.135, surdutovich2013biodamage}:
\begin{equation}
\sum_i^N \frac{1}{2} m_i \left( \alpha v_i \right)^2 = \frac{3 N k_{B} T}{2} + S_e \, l \ .
\label{eq:velocity_scaling}
\end{equation}
Here $S_e$ is the LET of the projectile ion, $l$ is the length of the simulation box in the $z$-direction (parallel to the ion's path), and $N$ is the total number of atoms within the ``hot'' cylinder. The first term on the right-hand side of Eq.~(\ref{eq:velocity_scaling}) is the kinetic energy of the 1-nm radius cylinder at the equilibrium temperature, $T = 300$~K, whereas the second term describes the energy loss by the ion as it propagates through the medium.
Note that non-uniform radial distribution of deposited energy, which accounts for the transport of energetic $\delta$-electrons, was addressed in an earlier study \cite{deVera_2017_EPJD.71.281}. The cited study demonstrated that the more accurate radial dose distribution (as compared to the uniform energy distribution within the cylinder of 1~nm radius) does not affect the shock wave dynamics for ions in the Bragg peak region in any significant way.

The shock wave propagates in the molecular system radially away from the ion track, see Fig.~\ref{fig:MSA_SW_DNA-density}A. The range of the shock wave propagation in the aqueous environment can be determined by monitoring the radial density of the water molecules in time. The ion track was oriented in this case directly through the geometrical center of the DNA strand. The results of this analysis are shown in Fig.~\ref{fig:MSA_SW_DNA-density}B. The wave front moves toward the edge of the simulation box as time passes, and the wave profile becomes lower and broader, showing that the shock wave relaxes as time passes. This indicates that the impact of the shock wave weakens over time. Figure~\ref{fig:MSA_SW_DNA-density}B shows that the maximal density of the wave does not change significantly in the range of $2 - 6$~nm from the ion track, thus a DNA strand placed in this range is expected to receive the strongest impact from the shock wave.

\begin{figure}[t!]
    \centering
    \includegraphics[width=0.45\textwidth]{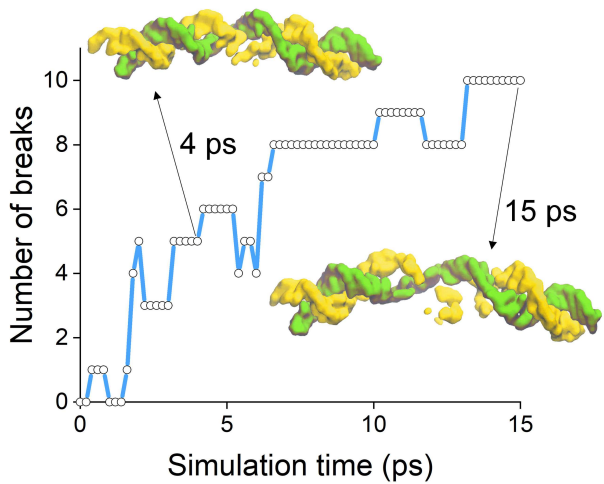}
    \caption{The number of strand breaks in the 30 base pairs-long DNA segment induced by the thermomechanical stress by the argon ion with an LET of 2890~eV/nm as a function of simulation time \cite{Friis2020}. Two insets in the top show the breakage of the DNA strands at simulation time instances of 4~ps and 15~ps. }
    \label{fig:MSA_SW_bond-breaks}
\end{figure}

The impact of a shock wave on DNA can be characterized through the probability of strand breaks formation, which can be used to quantify the amount of biodamage induced by the shock wave mechanism \cite{Surdutovich_2014_EPJD.68.353}.
Figure~\ref{fig:MSA_SW_bond-breaks} shows the number of strand breaks in the DNA segment caused by a shock wave induced by an argon ion with $S_e = 2890$~eV/nm, as a function of simulation time. The number of breaks rises quickly within the first 7~ps of the simulation, where the shock wave front hits both DNA strands of the target molecule. After this moment the DNA damage rate becomes lower lasting until approximately 15 ps, hereafter the number of breaks remains steady.
Even if the number of breaks generally rises with time, some fluctuations of this quantity can be seen locally. This effect might be attributed to the broken bonds which can be rejoined if the atoms involved get close to each other after the initial bond breakage. This analysis shows that multiple or even complex stand breaks might be induced by the generated shock waves.

The methodology reviewed in this Section can be applied in further computational studies considering irradiation of DNA with different ions and different orientations between the ion's path and the DNA molecule. Such analysis is important for understanding the radiation damage with ions on a quantitative level, focusing on particular physical, chemical, and biological effects that bring about lethal damage to cells exposed to ion beams \cite{Surdutovich_2014_EPJD.68.353, verkhovtsev2016multiscale, verkhovtsev2019phenomenon}.

\section{Conclusions}

\begin{sloppypar}
\MBNExplorer and \MBNStudio are powerful tools for computational modeling in different areas of challenging research arising in connection with the development of the novel and emerging technologies.
Illustrative case studies of multiscale modeling have been presented in this paper in relation to three emerging technologies, namely (i) development of novel sources of monochromatic high-energy radiation based on the crystalline undulators, (ii) controlled fabrication of nanostructures using the focused electron-beam induced deposition, and (iii) ion-beam cancer therapy.
These examples illustrate that the unique algorithms and methodologies implemented in \MBNExplorer, combined with the visualization interface and additional tools of \MBNStudio, in many cases can substitute expensive laboratory experiments by computational multiscale modeling, making the software play a role of a ``computational nano- and microscope''.
\end{sloppypar}

\section*{Acknowledgements}

\begin{sloppypar}
The authors are grateful for financial support from the COST Action CA17126 ``Towards understanding and modelling intense electronic excitation'' (TUMIEE), Deutsche Forschungsgemeinschaft (Projects no. 415716638, 413220201, SFB1372 and GRK1885), the Volkswagen Stiftung (Lichtenberg professorship to IAS), and the European Union's Horizon 2020 research and innovation programme -- the Radio-NP project (GA 794733) within the H2020-MSCA-IF-2017 call, the RADON project (GA 872494) within the H2020-MSCA-RISE-2019 call and the N-LIGHT project within the H2020-MSCA-RISE-2019 call (GA 872196).
\end{sloppypar}

\bibliography{bibliography}

\begin{thebibliography}{100}%
\makeatletter
\providecommand \@ifxundefined [1]{%
 \@ifx{#1\undefined}
}%
\providecommand \@ifnum [1]{%
 \ifnum #1\expandafter \@firstoftwo
 \else \expandafter \@secondoftwo
 \fi
}%
\providecommand \@ifx [1]{%
 \ifx #1\expandafter \@firstoftwo
 \else \expandafter \@secondoftwo
 \fi
}%
\providecommand \natexlab [1]{#1}%
\providecommand \enquote  [1]{``#1''}%
\providecommand \bibnamefont  [1]{#1}%
\providecommand \bibfnamefont [1]{#1}%
\providecommand \citenamefont [1]{#1}%
\providecommand \href@noop [0]{\@secondoftwo}%
\providecommand \href [0]{\begingroup \@sanitize@url \@href}%
\providecommand \@href[1]{\@@startlink{#1}\@@href}%
\providecommand \@@href[1]{\endgroup#1\@@endlink}%
\providecommand \@sanitize@url [0]{\catcode `\\12\catcode `\$12\catcode
  `\&12\catcode `\#12\catcode `\^12\catcode `\_12\catcode `\%12\relax}%
\providecommand \@@startlink[1]{}%
\providecommand \@@endlink[0]{}%
\providecommand \url  [0]{\begingroup\@sanitize@url \@url }%
\providecommand \@url [1]{\endgroup\@href {#1}{\urlprefix }}%
\providecommand \urlprefix  [0]{URL }%
\providecommand \Eprint [0]{\href }%
\providecommand \doibase [0]{https://doi.org/}%
\providecommand \selectlanguage [0]{\@gobble}%
\providecommand \bibinfo  [0]{\@secondoftwo}%
\providecommand \bibfield  [0]{\@secondoftwo}%
\providecommand \translation [1]{[#1]}%
\providecommand \BibitemOpen [0]{}%
\providecommand \bibitemStop [0]{}%
\providecommand \bibitemNoStop [0]{.\EOS\space}%
\providecommand \EOS [0]{\spacefactor3000\relax}%
\providecommand \BibitemShut  [1]{\csname bibitem#1\endcsname}%
\let\auto@bib@innerbib\@empty
\bibitem [{\citenamefont {Solov'yov}\ \emph
  {et~al.}(2017{\natexlab{a}})\citenamefont {Solov'yov}, \citenamefont
  {Korol},\ and\ \citenamefont {Solov'yov}}]{Solov'yov2017}%
  \BibitemOpen
  \bibfield  {author} {\bibinfo {author} {\bibfnamefont {I.~A.}\ \bibnamefont
  {Solov'yov}}, \bibinfo {author} {\bibfnamefont {A.~V.}\ \bibnamefont
  {Korol}},\ and\ \bibinfo {author} {\bibfnamefont {A.~V.}\ \bibnamefont
  {Solov'yov}},\ }\href@noop {} {\emph {\bibinfo {title} {{Multiscale Modeling
  of Complex Molecular Structure and Dynamics with MBN Explorer }}}}\ (\bibinfo
   {publisher} {Springer International Publishing},\ \bibinfo {address} {Cham,
  Switzerland},\ \bibinfo {year} {2017})\BibitemShut {NoStop}%
\bibitem [{\citenamefont {Solov'yov}\ \emph {et~al.}(2012)\citenamefont
  {Solov'yov}, \citenamefont {Yakubovich}, \citenamefont {Nikolaev},
  \citenamefont {Volkovets},\ and\ \citenamefont {Solov'yov}}]{Solovyov2012}%
  \BibitemOpen
  \bibfield  {author} {\bibinfo {author} {\bibfnamefont {I.~A.}\ \bibnamefont
  {Solov'yov}}, \bibinfo {author} {\bibfnamefont {A.~V.}\ \bibnamefont
  {Yakubovich}}, \bibinfo {author} {\bibfnamefont {P.~V.}\ \bibnamefont
  {Nikolaev}}, \bibinfo {author} {\bibfnamefont {I.}~\bibnamefont
  {Volkovets}},\ and\ \bibinfo {author} {\bibfnamefont {A.~V.}\ \bibnamefont
  {Solov'yov}},\ }\href@noop {} {\bibfield  {journal} {\bibinfo  {journal} {J.
  Comput. Chem.}\ }\textbf {\bibinfo {volume} {33}},\ \bibinfo {pages} {2412}
  (\bibinfo {year} {2012})}\BibitemShut {NoStop}%
\bibitem [{\citenamefont {Sushko}\ \emph {et~al.}(2019)\citenamefont {Sushko},
  \citenamefont {Solov'yov},\ and\ \citenamefont {Solov'yov}}]{Sushko2019}%
  \BibitemOpen
  \bibfield  {author} {\bibinfo {author} {\bibfnamefont {G.}~\bibnamefont
  {Sushko}}, \bibinfo {author} {\bibfnamefont {I.}~\bibnamefont {Solov'yov}},\
  and\ \bibinfo {author} {\bibfnamefont {A.}~\bibnamefont {Solov'yov}},\
  }\href@noop {} {\bibfield  {journal} {\bibinfo  {journal} {J. Mol. Graph.
  Model.}\ }\textbf {\bibinfo {volume} {88}},\ \bibinfo {pages} {247} (\bibinfo
  {year} {2019})}\BibitemShut {NoStop}%
\bibitem [{\citenamefont {Connerade}\ and\ \citenamefont
  {Solov'yov}(2008)}]{ISACC_book_2008}%
  \BibitemOpen
  \bibinfo {editor} {\bibfnamefont {J.-P.}\ \bibnamefont {Connerade}}\ and\
  \bibinfo {editor} {\bibfnamefont {A.~V.}\ \bibnamefont {Solov'yov}},\ eds.,\
  \href@noop {} {\emph {\bibinfo {title} {Latest Advances in Atomic Clusters
  Collision: Structure and Dynamics from the Nuclear to the Biological
  Scale}}}\ (\bibinfo  {publisher} {Imperial College Press},\ \bibinfo {year}
  {2008})\BibitemShut {NoStop}%
\bibitem [{\citenamefont {Senn}\ and\ \citenamefont
  {Thiel}(2009)}]{Senn_2009_AngChemIntEd.48.1198}%
  \BibitemOpen
  \bibfield  {author} {\bibinfo {author} {\bibfnamefont {H.~M.}\ \bibnamefont
  {Senn}}\ and\ \bibinfo {author} {\bibfnamefont {W.}~\bibnamefont {Thiel}},\
  }\href@noop {} {\bibfield  {journal} {\bibinfo  {journal} {Angew. Chem. Int.
  Ed.}\ }\textbf {\bibinfo {volume} {48}},\ \bibinfo {pages} {1198} (\bibinfo
  {year} {2009})}\BibitemShut {NoStop}%
\bibitem [{\citenamefont {Chung}\ \emph {et~al.}(2015)\citenamefont {Chung},
  \citenamefont {Sameera}, \citenamefont {Ramozzi}, \citenamefont {Page},
  \citenamefont {Hatanaka}, \citenamefont {Petrova}, \citenamefont {Harris},
  \citenamefont {Li}, \citenamefont {Ke}, \citenamefont {Liu}, \citenamefont
  {Li}, \citenamefont {Ding},\ and\ \citenamefont
  {Morokuma}}]{Chung_2015_ChemRev.115.5678}%
  \BibitemOpen
  \bibfield  {author} {\bibinfo {author} {\bibfnamefont {L.~W.}\ \bibnamefont
  {Chung}}, \bibinfo {author} {\bibfnamefont {W.~M.~C.}\ \bibnamefont
  {Sameera}}, \bibinfo {author} {\bibfnamefont {R.}~\bibnamefont {Ramozzi}},
  \bibinfo {author} {\bibfnamefont {A.~J.}\ \bibnamefont {Page}}, \bibinfo
  {author} {\bibfnamefont {M.}~\bibnamefont {Hatanaka}}, \bibinfo {author}
  {\bibfnamefont {G.~P.}\ \bibnamefont {Petrova}}, \bibinfo {author}
  {\bibfnamefont {T.~V.}\ \bibnamefont {Harris}}, \bibinfo {author}
  {\bibfnamefont {X.}~\bibnamefont {Li}}, \bibinfo {author} {\bibfnamefont
  {Z.}~\bibnamefont {Ke}}, \bibinfo {author} {\bibfnamefont {F.}~\bibnamefont
  {Liu}}, \bibinfo {author} {\bibfnamefont {H.-B.}\ \bibnamefont {Li}},
  \bibinfo {author} {\bibfnamefont {L.}~\bibnamefont {Ding}},\ and\ \bibinfo
  {author} {\bibfnamefont {K.}~\bibnamefont {Morokuma}},\ }\href@noop {}
  {\bibfield  {journal} {\bibinfo  {journal} {Chem. Rev.}\ }\textbf {\bibinfo
  {volume} {115}},\ \bibinfo {pages} {5678} (\bibinfo {year}
  {2015})}\BibitemShut {NoStop}%
\bibitem [{\citenamefont {Brunk}\ and\ \citenamefont
  {Rothlisberger}(2015)}]{Brunk_2015_ChemRev.115.6217}%
  \BibitemOpen
  \bibfield  {author} {\bibinfo {author} {\bibfnamefont {E.}~\bibnamefont
  {Brunk}}\ and\ \bibinfo {author} {\bibfnamefont {U.}~\bibnamefont
  {Rothlisberger}},\ }\href@noop {} {\bibfield  {journal} {\bibinfo  {journal}
  {Chem. Rev.}\ }\textbf {\bibinfo {volume} {115}},\ \bibinfo {pages} {6217}
  (\bibinfo {year} {2015})}\BibitemShut {NoStop}%
\bibitem [{\citenamefont {Sanbonmatsu}\ and\ \citenamefont
  {Tung}(2007)}]{Sanbonmatsu2007}%
  \BibitemOpen
  \bibfield  {author} {\bibinfo {author} {\bibfnamefont {K.~Y.}\ \bibnamefont
  {Sanbonmatsu}}\ and\ \bibinfo {author} {\bibfnamefont {C.-S.}\ \bibnamefont
  {Tung}},\ }\href@noop {} {\bibfield  {journal} {\bibinfo  {journal} {J.
  Struct. Biol.}\ }\textbf {\bibinfo {volume} {157}},\ \bibinfo {pages} {470}
  (\bibinfo {year} {2007})}\BibitemShut {NoStop}%
\bibitem [{\citenamefont {Zhao}\ \emph {et~al.}(2013)\citenamefont {Zhao},
  \citenamefont {Perilla}, \citenamefont {Yufenyuy}, \citenamefont {Meng},
  \citenamefont {Chen}, \citenamefont {Ning}, \citenamefont {Ahn},
  \citenamefont {Gronenborn}, \citenamefont {Schulten}, \citenamefont {Aiken}
  \emph {et~al.}}]{Zhao2013}%
  \BibitemOpen
  \bibfield  {author} {\bibinfo {author} {\bibfnamefont {G.}~\bibnamefont
  {Zhao}}, \bibinfo {author} {\bibfnamefont {J.~R.}\ \bibnamefont {Perilla}},
  \bibinfo {author} {\bibfnamefont {E.~L.}\ \bibnamefont {Yufenyuy}}, \bibinfo
  {author} {\bibfnamefont {X.}~\bibnamefont {Meng}}, \bibinfo {author}
  {\bibfnamefont {B.}~\bibnamefont {Chen}}, \bibinfo {author} {\bibfnamefont
  {J.}~\bibnamefont {Ning}}, \bibinfo {author} {\bibfnamefont {J.}~\bibnamefont
  {Ahn}}, \bibinfo {author} {\bibfnamefont {A.~M.}\ \bibnamefont {Gronenborn}},
  \bibinfo {author} {\bibfnamefont {K.}~\bibnamefont {Schulten}}, \bibinfo
  {author} {\bibfnamefont {C.}~\bibnamefont {Aiken}}, \emph {et~al.},\
  }\href@noop {} {\bibfield  {journal} {\bibinfo  {journal} {Nature}\ }\textbf
  {\bibinfo {volume} {497}},\ \bibinfo {pages} {643} (\bibinfo {year}
  {2013})}\BibitemShut {NoStop}%
\bibitem [{\citenamefont {Schulz}\ \emph {et~al.}(2009)\citenamefont {Schulz},
  \citenamefont {Lindner}, \citenamefont {Petridis},\ and\ \citenamefont
  {Smith}}]{Schulz_2009_JCTC.5.2798}%
  \BibitemOpen
  \bibfield  {author} {\bibinfo {author} {\bibfnamefont {R.}~\bibnamefont
  {Schulz}}, \bibinfo {author} {\bibfnamefont {B.}~\bibnamefont {Lindner}},
  \bibinfo {author} {\bibfnamefont {L.}~\bibnamefont {Petridis}},\ and\
  \bibinfo {author} {\bibfnamefont {J.~C.}\ \bibnamefont {Smith}},\ }\href@noop
  {} {\bibfield  {journal} {\bibinfo  {journal} {J. Chem. Theory Comput.}\
  }\textbf {\bibinfo {volume} {5}},\ \bibinfo {pages} {2798} (\bibinfo {year}
  {2009})}\BibitemShut {NoStop}%
\bibitem [{\citenamefont {Bouda{\"i}ffa}\ \emph {et~al.}(2000)\citenamefont
  {Bouda{\"i}ffa}, \citenamefont {Cloutier}, \citenamefont {Hunting},
  \citenamefont {Huels},\ and\ \citenamefont {Sanche}}]{DNA2}%
  \BibitemOpen
  \bibfield  {author} {\bibinfo {author} {\bibfnamefont {B.}~\bibnamefont
  {Bouda{\"i}ffa}}, \bibinfo {author} {\bibfnamefont {P.}~\bibnamefont
  {Cloutier}}, \bibinfo {author} {\bibfnamefont {D.}~\bibnamefont {Hunting}},
  \bibinfo {author} {\bibfnamefont {M.~A.}\ \bibnamefont {Huels}},\ and\
  \bibinfo {author} {\bibfnamefont {L.}~\bibnamefont {Sanche}},\ }\href@noop {}
  {\bibfield  {journal} {\bibinfo  {journal} {Science}\ }\textbf {\bibinfo
  {volume} {287}},\ \bibinfo {pages} {1658} (\bibinfo {year}
  {2000})}\BibitemShut {NoStop}%
\bibitem [{\citenamefont {Huels}\ \emph {et~al.}(2003)\citenamefont {Huels},
  \citenamefont {Bouda{\"i}ffa}, \citenamefont {Cloutier}, \citenamefont
  {Hunting},\ and\ \citenamefont {Sanche}}]{DNA3}%
  \BibitemOpen
  \bibfield  {author} {\bibinfo {author} {\bibfnamefont {M.~A.}\ \bibnamefont
  {Huels}}, \bibinfo {author} {\bibfnamefont {B.}~\bibnamefont
  {Bouda{\"i}ffa}}, \bibinfo {author} {\bibfnamefont {P.}~\bibnamefont
  {Cloutier}}, \bibinfo {author} {\bibfnamefont {D.}~\bibnamefont {Hunting}},\
  and\ \bibinfo {author} {\bibfnamefont {L.}~\bibnamefont {Sanche}},\
  }\href@noop {} {\bibfield  {journal} {\bibinfo  {journal} {J. Am. Chem.
  Soc.}\ }\textbf {\bibinfo {volume} {125}},\ \bibinfo {pages} {4467} (\bibinfo
  {year} {2003})}\BibitemShut {NoStop}%
\bibitem [{\citenamefont {Surdutovich}\ and\ \citenamefont
  {Solov'yov}(2014)}]{Surdutovich_2014_EPJD.68.353}%
  \BibitemOpen
  \bibfield  {author} {\bibinfo {author} {\bibfnamefont {E.}~\bibnamefont
  {Surdutovich}}\ and\ \bibinfo {author} {\bibfnamefont {A.~V.}\ \bibnamefont
  {Solov'yov}},\ }\href@noop {} {\bibfield  {journal} {\bibinfo  {journal}
  {Eur. Phys. J. D}\ }\textbf {\bibinfo {volume} {68}},\ \bibinfo {pages} {353}
  (\bibinfo {year} {2014})}\BibitemShut {NoStop}%
\bibitem [{\citenamefont {Haume}\ \emph {et~al.}(2016)\citenamefont {Haume},
  \citenamefont {Rosa}, \citenamefont {Grellet}, \citenamefont
  {\'{S}mia{\l}ek}, \citenamefont {Butterworth}, \citenamefont {Solov'yov},
  \citenamefont {Prise}, \citenamefont {Golding},\ and\ \citenamefont
  {Mason}}]{CNano_GNPs_review}%
  \BibitemOpen
  \bibfield  {author} {\bibinfo {author} {\bibfnamefont {K.}~\bibnamefont
  {Haume}}, \bibinfo {author} {\bibfnamefont {S.}~\bibnamefont {Rosa}},
  \bibinfo {author} {\bibfnamefont {S.}~\bibnamefont {Grellet}}, \bibinfo
  {author} {\bibfnamefont {M.~A.}\ \bibnamefont {\'{S}mia{\l}ek}}, \bibinfo
  {author} {\bibfnamefont {K.~T.}\ \bibnamefont {Butterworth}}, \bibinfo
  {author} {\bibfnamefont {A.~V.}\ \bibnamefont {Solov'yov}}, \bibinfo {author}
  {\bibfnamefont {K.~M.}\ \bibnamefont {Prise}}, \bibinfo {author}
  {\bibfnamefont {J.}~\bibnamefont {Golding}},\ and\ \bibinfo {author}
  {\bibfnamefont {N.~J.}\ \bibnamefont {Mason}},\ }\href@noop {} {\bibfield
  {journal} {\bibinfo  {journal} {Cancer Nanotechnol.}\ }\textbf {\bibinfo
  {volume} {7}},\ \bibinfo {pages} {8} (\bibinfo {year} {2016})}\BibitemShut
  {NoStop}%
\bibitem [{\citenamefont {Kmiecik}\ \emph {et~al.}(2016)\citenamefont
  {Kmiecik}, \citenamefont {Gront}, \citenamefont {Kolinski}, \citenamefont
  {Wieteska}, \citenamefont {Dawid},\ and\ \citenamefont
  {Kolinski}}]{Kmiecik_2016_ChemRev.116.7898}%
  \BibitemOpen
  \bibfield  {author} {\bibinfo {author} {\bibfnamefont {S.}~\bibnamefont
  {Kmiecik}}, \bibinfo {author} {\bibfnamefont {D.}~\bibnamefont {Gront}},
  \bibinfo {author} {\bibfnamefont {M.}~\bibnamefont {Kolinski}}, \bibinfo
  {author} {\bibfnamefont {L.}~\bibnamefont {Wieteska}}, \bibinfo {author}
  {\bibfnamefont {A.~E.}\ \bibnamefont {Dawid}},\ and\ \bibinfo {author}
  {\bibfnamefont {A.}~\bibnamefont {Kolinski}},\ }\href@noop {} {\bibfield
  {journal} {\bibinfo  {journal} {Chem. Rev.}\ }\textbf {\bibinfo {volume}
  {116}},\ \bibinfo {pages} {7898} (\bibinfo {year} {2016})}\BibitemShut
  {NoStop}%
\bibitem [{\citenamefont {Dick}\ \emph {et~al.}(2011)\citenamefont {Dick},
  \citenamefont {Solov'yov},\ and\ \citenamefont {Solov'yov}}]{Dick2011}%
  \BibitemOpen
  \bibfield  {author} {\bibinfo {author} {\bibfnamefont {V.~V.}\ \bibnamefont
  {Dick}}, \bibinfo {author} {\bibfnamefont {I.~A.}\ \bibnamefont
  {Solov'yov}},\ and\ \bibinfo {author} {\bibfnamefont {A.~V.}\ \bibnamefont
  {Solov'yov}},\ }\href@noop {} {\bibfield  {journal} {\bibinfo  {journal}
  {Phys. Rev. B}\ }\textbf {\bibinfo {volume} {84}},\ \bibinfo {pages} {115408}
  (\bibinfo {year} {2011})}\BibitemShut {NoStop}%
\bibitem [{\citenamefont {Panshenskov}\ \emph {et~al.}(2014)\citenamefont
  {Panshenskov}, \citenamefont {Solov'yov},\ and\ \citenamefont
  {Solov'yov}}]{3DKMC}%
  \BibitemOpen
  \bibfield  {author} {\bibinfo {author} {\bibfnamefont {M.~A.}\ \bibnamefont
  {Panshenskov}}, \bibinfo {author} {\bibfnamefont {I.~A.}\ \bibnamefont
  {Solov'yov}},\ and\ \bibinfo {author} {\bibfnamefont {A.~V.}\ \bibnamefont
  {Solov'yov}},\ }\href@noop {} {\bibfield  {journal} {\bibinfo  {journal} {J.
  Comput. Chem.}\ }\textbf {\bibinfo {volume} {35}},\ \bibinfo {pages} {1317}
  (\bibinfo {year} {2014})}\BibitemShut {NoStop}%
\bibitem [{\citenamefont {Moskovkin}\ \emph {et~al.}(2014)\citenamefont
  {Moskovkin}, \citenamefont {Panshenskov}, \citenamefont {Lucas},\ and\
  \citenamefont {Solov'yov}}]{Moskovkin_2014_PSSB.251.1456}%
  \BibitemOpen
  \bibfield  {author} {\bibinfo {author} {\bibfnamefont {P.}~\bibnamefont
  {Moskovkin}}, \bibinfo {author} {\bibfnamefont {M.~A.}\ \bibnamefont
  {Panshenskov}}, \bibinfo {author} {\bibfnamefont {S.}~\bibnamefont {Lucas}},\
  and\ \bibinfo {author} {\bibfnamefont {A.~V.}\ \bibnamefont {Solov'yov}},\
  }\href@noop {} {\bibfield  {journal} {\bibinfo  {journal} {Phys. Stat. Sol.
  B}\ }\textbf {\bibinfo {volume} {251}},\ \bibinfo {pages} {1456} (\bibinfo
  {year} {2014})}\BibitemShut {NoStop}%
\bibitem [{\citenamefont {Logan}(2016)}]{Logan_FEM_book}%
  \BibitemOpen
  \bibfield  {author} {\bibinfo {author} {\bibfnamefont {D.~L.}\ \bibnamefont
  {Logan}},\ }\href@noop {} {\emph {\bibinfo {title} {A First Course in the
  Finite Element Method (6th ed.)}}}\ (\bibinfo  {publisher} {Cengage
  Learning},\ \bibinfo {address} {Boston, MA},\ \bibinfo {year}
  {2016})\BibitemShut {NoStop}%
\bibitem [{\citenamefont {Solov'yov}\ \emph
  {et~al.}(2017{\natexlab{b}})\citenamefont {Solov'yov}, \citenamefont
  {Sushko},\ and\ \citenamefont {Solov'yov}}]{Solo2017}%
  \BibitemOpen
  \bibfield  {author} {\bibinfo {author} {\bibfnamefont {I.~A.}\ \bibnamefont
  {Solov'yov}}, \bibinfo {author} {\bibfnamefont {G.}~\bibnamefont {Sushko}},\
  and\ \bibinfo {author} {\bibfnamefont {A.~V.}\ \bibnamefont {Solov'yov}},\
  }\href@noop {} {\emph {\bibinfo {title} {{MBN Explorer Users' Guide. Version
  3.0}}}}\ (\bibinfo  {publisher} {MesoBioNano Science Publishing},\ \bibinfo
  {address} {Frankfurt am Main},\ \bibinfo {year} {2017})\BibitemShut {NoStop}%
\bibitem [{\citenamefont {Solov'yov}\ \emph
  {et~al.}(2017{\natexlab{c}})\citenamefont {Solov'yov}, \citenamefont
  {Sushko}, \citenamefont {Verkhovtsev}, \citenamefont {Korol},\ and\
  \citenamefont {Solov'yov}}]{Solov2017}%
  \BibitemOpen
  \bibfield  {author} {\bibinfo {author} {\bibfnamefont {I.~A.}\ \bibnamefont
  {Solov'yov}}, \bibinfo {author} {\bibfnamefont {G.}~\bibnamefont {Sushko}},
  \bibinfo {author} {\bibfnamefont {A.}~\bibnamefont {Verkhovtsev}}, \bibinfo
  {author} {\bibfnamefont {A.~V.}\ \bibnamefont {Korol}},\ and\ \bibinfo
  {author} {\bibfnamefont {A.~V.}\ \bibnamefont {Solov'yov}},\ }\href@noop {}
  {\emph {\bibinfo {title} {{MBN Explorer and MBN Studio Tutorials: Version
  3.0}}}}\ (\bibinfo  {publisher} {MesoBioNano Science Publishing},\ \bibinfo
  {address} {Frankfurt am Main},\ \bibinfo {year} {2017})\ p.\ \bibinfo {pages}
  {288}\BibitemShut {NoStop}%
\bibitem [{\citenamefont {Korol}\ and\ \citenamefont
  {Solov'yov}(2020)}]{Korol2020_LS_review}%
  \BibitemOpen
  \bibfield  {author} {\bibinfo {author} {\bibfnamefont {A.~V.}\ \bibnamefont
  {Korol}}\ and\ \bibinfo {author} {\bibfnamefont {A.~V.}\ \bibnamefont
  {Solov'yov}},\ }\href@noop {} {\bibfield  {journal} {\bibinfo  {journal}
  {Eur. Phys. J. D}\ }\textbf {\bibinfo {volume} {74}},\ \bibinfo {pages} {201}
  (\bibinfo {year} {2020})}\BibitemShut {NoStop}%
\bibitem [{\citenamefont {Korol}\ \emph {et~al.}(2014)\citenamefont {Korol},
  \citenamefont {Solov'yov},\ and\ \citenamefont {Greiner}}]{Channeling_book}%
  \BibitemOpen
  \bibfield  {author} {\bibinfo {author} {\bibfnamefont {A.~V.}\ \bibnamefont
  {Korol}}, \bibinfo {author} {\bibfnamefont {A.~V.}\ \bibnamefont
  {Solov'yov}},\ and\ \bibinfo {author} {\bibfnamefont {W.}~\bibnamefont
  {Greiner}},\ }\href@noop {} {\emph {\bibinfo {title} {Channeling and
  Radiation in Periodically Bent Crystals (2nd ed.)}}}\ (\bibinfo  {publisher}
  {Springer Series on Atomic, Optical, and Plasma Physics, vol. 69.
  Springer-Verlag},\ \bibinfo {address} {Heidelberg, New York, Dordrecht,
  London},\ \bibinfo {year} {2014})\ p.\ \bibinfo {pages} {284}\BibitemShut
  {NoStop}%
\bibitem [{\citenamefont {Korol}\ \emph {et~al.}(2021)\citenamefont {Korol},
  \citenamefont {Sushko},\ and\ \citenamefont
  {Solov'yov}}]{Korol2021_EPJD_Collo}%
  \BibitemOpen
  \bibfield  {author} {\bibinfo {author} {\bibfnamefont {A.~V.}\ \bibnamefont
  {Korol}}, \bibinfo {author} {\bibfnamefont {G.~B.}\ \bibnamefont {Sushko}},\
  and\ \bibinfo {author} {\bibfnamefont {A.~V.}\ \bibnamefont {Solov'yov}},\
  }\href@noop {} {\bibfield  {journal} {\bibinfo  {journal} {Eur. Phys. J. D}\
  }\textbf {\bibinfo {volume} {75}},\ \bibinfo {pages} {107} (\bibinfo {year}
  {2021})}\BibitemShut {NoStop}%
\bibitem [{\citenamefont {van Dorp}\ and\ \citenamefont
  {Hagen}(2008)}]{VanDorp2008}%
  \BibitemOpen
  \bibfield  {author} {\bibinfo {author} {\bibfnamefont {W.~F.}\ \bibnamefont
  {van Dorp}}\ and\ \bibinfo {author} {\bibfnamefont {C.~W.}\ \bibnamefont
  {Hagen}},\ }\href@noop {} {\bibfield  {journal} {\bibinfo  {journal} {J.
  Appl. Phys.}\ }\textbf {\bibinfo {volume} {104}},\ \bibinfo {pages} {081301}
  (\bibinfo {year} {2008})}\BibitemShut {NoStop}%
\bibitem [{\citenamefont {Utke}\ \emph {et~al.}(2012)\citenamefont {Utke},
  \citenamefont {Moshkalev},\ and\ \citenamefont {Russel}}]{Utke_book_2012}%
  \BibitemOpen
  \bibinfo {editor} {\bibfnamefont {I.}~\bibnamefont {Utke}}, \bibinfo {editor}
  {\bibfnamefont {S.}~\bibnamefont {Moshkalev}},\ and\ \bibinfo {editor}
  {\bibfnamefont {P.}~\bibnamefont {Russel}},\ eds.,\ \href@noop {} {\emph
  {\bibinfo {title} {{Nanofabrication Using Focused Ion and Electron Beams}}}}\
  (\bibinfo  {publisher} {Oxford University Press},\ \bibinfo {year}
  {2012})\BibitemShut {NoStop}%
\bibitem [{\citenamefont {Huth}\ \emph {et~al.}(2012)\citenamefont {Huth},
  \citenamefont {Porrati}, \citenamefont {Schwalb}, \citenamefont {Winhold},
  \citenamefont {Sachser}, \citenamefont {Dukic}, \citenamefont {Adams},\ and\
  \citenamefont {Fantner}}]{Huth2012}%
  \BibitemOpen
  \bibfield  {author} {\bibinfo {author} {\bibfnamefont {M.}~\bibnamefont
  {Huth}}, \bibinfo {author} {\bibfnamefont {F.}~\bibnamefont {Porrati}},
  \bibinfo {author} {\bibfnamefont {C.}~\bibnamefont {Schwalb}}, \bibinfo
  {author} {\bibfnamefont {M.}~\bibnamefont {Winhold}}, \bibinfo {author}
  {\bibfnamefont {R.}~\bibnamefont {Sachser}}, \bibinfo {author} {\bibfnamefont
  {M.}~\bibnamefont {Dukic}}, \bibinfo {author} {\bibfnamefont
  {J.}~\bibnamefont {Adams}},\ and\ \bibinfo {author} {\bibfnamefont
  {G.}~\bibnamefont {Fantner}},\ }\href@noop {} {\bibfield  {journal} {\bibinfo
   {journal} {Beilstein J. Nanotechnol.}\ }\textbf {\bibinfo {volume} {3}},\
  \bibinfo {pages} {597} (\bibinfo {year} {2012})}\BibitemShut {NoStop}%
\bibitem [{\citenamefont {Schardt}\ \emph {et~al.}(2010)\citenamefont
  {Schardt}, \citenamefont {Els{\"a}sser},\ and\ \citenamefont
  {Schulz-Ertner}}]{schardt2010heavy}%
  \BibitemOpen
  \bibfield  {author} {\bibinfo {author} {\bibfnamefont {D.}~\bibnamefont
  {Schardt}}, \bibinfo {author} {\bibfnamefont {T.}~\bibnamefont
  {Els{\"a}sser}},\ and\ \bibinfo {author} {\bibfnamefont {D.}~\bibnamefont
  {Schulz-Ertner}},\ }\href@noop {} {\bibfield  {journal} {\bibinfo  {journal}
  {Rev. Mod. Phys.}\ }\textbf {\bibinfo {volume} {82}},\ \bibinfo {pages} {383}
  (\bibinfo {year} {2010})}\BibitemShut {NoStop}%
\bibitem [{\citenamefont {Solov'yov}(2017)}]{solov2016nanoscale}%
  \BibitemOpen
  \bibinfo {editor} {\bibfnamefont {A.~V.}\ \bibnamefont {Solov'yov}},\ ed.,\
  \href@noop {} {\emph {\bibinfo {title} {Nanoscale Insights into Ion-Beam
  Cancer Therapy}}}\ (\bibinfo  {publisher} {Springer International
  Publishing},\ \bibinfo {address} {Cham, Switzerland},\ \bibinfo {year}
  {2017})\BibitemShut {NoStop}%
\bibitem [{\citenamefont {Korol}\ \emph {et~al.}(1998)\citenamefont {Korol},
  \citenamefont {Solov'yov},\ and\ \citenamefont {Greiner}}]{KSG1998}%
  \BibitemOpen
  \bibfield  {author} {\bibinfo {author} {\bibfnamefont {A.~V.}\ \bibnamefont
  {Korol}}, \bibinfo {author} {\bibfnamefont {A.~V.}\ \bibnamefont
  {Solov'yov}},\ and\ \bibinfo {author} {\bibfnamefont {W.}~\bibnamefont
  {Greiner}},\ }\href@noop {} {\bibfield  {journal} {\bibinfo  {journal} {J.
  Phys. G: Nucl. Part. Phys.}\ }\textbf {\bibinfo {volume} {24}},\ \bibinfo
  {pages} {L45} (\bibinfo {year} {1998})}\BibitemShut {NoStop}%
\bibitem [{\citenamefont {Korol}\ \emph {et~al.}(1999)\citenamefont {Korol},
  \citenamefont {Solov'yov},\ and\ \citenamefont {Greiner}}]{KSG_review_1999}%
  \BibitemOpen
  \bibfield  {author} {\bibinfo {author} {\bibfnamefont {A.~V.}\ \bibnamefont
  {Korol}}, \bibinfo {author} {\bibfnamefont {A.~V.}\ \bibnamefont
  {Solov'yov}},\ and\ \bibinfo {author} {\bibfnamefont {W.}~\bibnamefont
  {Greiner}},\ }\href@noop {} {\bibfield  {journal} {\bibinfo  {journal} {Int.
  J. Mod. Phys. E}\ }\textbf {\bibinfo {volume} {8}},\ \bibinfo {pages} {49}
  (\bibinfo {year} {1999})}\BibitemShut {NoStop}%
\bibitem [{\citenamefont {Korol}\ \emph {et~al.}(2004)\citenamefont {Korol},
  \citenamefont {Solov'yov},\ and\ \citenamefont {Greiner}}]{KSG_review2004}%
  \BibitemOpen
  \bibfield  {author} {\bibinfo {author} {\bibfnamefont {A.~V.}\ \bibnamefont
  {Korol}}, \bibinfo {author} {\bibfnamefont {A.~V.}\ \bibnamefont
  {Solov'yov}},\ and\ \bibinfo {author} {\bibfnamefont {W.}~\bibnamefont
  {Greiner}},\ }\href@noop {} {\bibfield  {journal} {\bibinfo  {journal} {Int.
  J. Mod. Phys. E}\ }\textbf {\bibinfo {volume} {13}},\ \bibinfo {pages} {867}
  (\bibinfo {year} {2004})}\BibitemShut {NoStop}%
\bibitem [{\citenamefont {Sushko}\ \emph {et~al.}(2013)\citenamefont {Sushko},
  \citenamefont {Bezchastnov}, \citenamefont {Solov'yov}, \citenamefont
  {Korol}, \citenamefont {Greiner},\ and\ \citenamefont
  {Solov'yov}}]{MBNExplorer_Chan}%
  \BibitemOpen
  \bibfield  {author} {\bibinfo {author} {\bibfnamefont {G.~B.}\ \bibnamefont
  {Sushko}}, \bibinfo {author} {\bibfnamefont {V.~G.}\ \bibnamefont
  {Bezchastnov}}, \bibinfo {author} {\bibfnamefont {I.~A.}\ \bibnamefont
  {Solov'yov}}, \bibinfo {author} {\bibfnamefont {A.~V.}\ \bibnamefont
  {Korol}}, \bibinfo {author} {\bibfnamefont {W.}~\bibnamefont {Greiner}},\
  and\ \bibinfo {author} {\bibfnamefont {A.~V.}\ \bibnamefont {Solov'yov}},\
  }\href@noop {} {\bibfield  {journal} {\bibinfo  {journal} {J. Comput. Phys.}\
  }\textbf {\bibinfo {volume} {252}},\ \bibinfo {pages} {404} (\bibinfo {year}
  {2013})}\BibitemShut {NoStop}%
\bibitem [{\citenamefont {Emma}\ \emph {et~al.}(2010)\citenamefont {Emma},
  \citenamefont {Akre}, \citenamefont {Arthur}, \citenamefont {Bionta},
  \citenamefont {Bostedt}, \citenamefont {Bozek}, \citenamefont {Brachmann},
  \citenamefont {Bucksbaum}, \citenamefont {Coffee}, \citenamefont {Decker},
  \citenamefont {Ding}, \citenamefont {Dowell}, \citenamefont {Edstrom},
  \citenamefont {Fisher}, \citenamefont {Frisch}, \citenamefont {Gilevich},
  \citenamefont {Hastings}, \citenamefont {Hays}, \citenamefont {Hering},
  \citenamefont {Huang}, \citenamefont {Iverson}, \citenamefont {Loos},
  \citenamefont {Messerschmidt}, \citenamefont {Miahnahri}, \citenamefont
  {Moeller}, \citenamefont {Nuhn}, \citenamefont {Pile}, \citenamefont
  {Ratner}, \citenamefont {Rzepiela}, \citenamefont {Schultz}, \citenamefont
  {Smith}, \citenamefont {Stefan}, \citenamefont {Tompkins}, \citenamefont
  {Turner}, \citenamefont {Welch}, \citenamefont {White}, \citenamefont {Wu},
  \citenamefont {Yocky},\ and\ \citenamefont {J.}}]{LCLS2010}%
  \BibitemOpen
  \bibfield  {author} {\bibinfo {author} {\bibfnamefont {P.}~\bibnamefont
  {Emma}}, \bibinfo {author} {\bibfnamefont {R.}~\bibnamefont {Akre}}, \bibinfo
  {author} {\bibfnamefont {J.}~\bibnamefont {Arthur}}, \bibinfo {author}
  {\bibfnamefont {R.}~\bibnamefont {Bionta}}, \bibinfo {author} {\bibfnamefont
  {C.}~\bibnamefont {Bostedt}}, \bibinfo {author} {\bibfnamefont
  {J.}~\bibnamefont {Bozek}}, \bibinfo {author} {\bibfnamefont
  {A.}~\bibnamefont {Brachmann}}, \bibinfo {author} {\bibfnamefont
  {P.}~\bibnamefont {Bucksbaum}}, \bibinfo {author} {\bibfnamefont
  {R.}~\bibnamefont {Coffee}}, \bibinfo {author} {\bibfnamefont {F.-J.}\
  \bibnamefont {Decker}}, \bibinfo {author} {\bibfnamefont {Y.}~\bibnamefont
  {Ding}}, \bibinfo {author} {\bibfnamefont {D.}~\bibnamefont {Dowell}},
  \bibinfo {author} {\bibfnamefont {S.}~\bibnamefont {Edstrom}}, \bibinfo
  {author} {\bibfnamefont {A.}~\bibnamefont {Fisher}}, \bibinfo {author}
  {\bibfnamefont {J.}~\bibnamefont {Frisch}}, \bibinfo {author} {\bibfnamefont
  {S.}~\bibnamefont {Gilevich}}, \bibinfo {author} {\bibfnamefont
  {J.}~\bibnamefont {Hastings}}, \bibinfo {author} {\bibfnamefont
  {G.}~\bibnamefont {Hays}}, \bibinfo {author} {\bibfnamefont {P.}~\bibnamefont
  {Hering}}, \bibinfo {author} {\bibfnamefont {Z.}~\bibnamefont {Huang}},
  \bibinfo {author} {\bibfnamefont {R.}~\bibnamefont {Iverson}}, \bibinfo
  {author} {\bibfnamefont {H.}~\bibnamefont {Loos}}, \bibinfo {author}
  {\bibfnamefont {M.}~\bibnamefont {Messerschmidt}}, \bibinfo {author}
  {\bibfnamefont {A.}~\bibnamefont {Miahnahri}}, \bibinfo {author}
  {\bibfnamefont {S.}~\bibnamefont {Moeller}}, \bibinfo {author} {\bibfnamefont
  {H.-D.}\ \bibnamefont {Nuhn}}, \bibinfo {author} {\bibfnamefont
  {G.}~\bibnamefont {Pile}}, \bibinfo {author} {\bibfnamefont {D.}~\bibnamefont
  {Ratner}}, \bibinfo {author} {\bibfnamefont {J.}~\bibnamefont {Rzepiela}},
  \bibinfo {author} {\bibfnamefont {D.}~\bibnamefont {Schultz}}, \bibinfo
  {author} {\bibfnamefont {T.}~\bibnamefont {Smith}}, \bibinfo {author}
  {\bibfnamefont {P.}~\bibnamefont {Stefan}}, \bibinfo {author} {\bibfnamefont
  {H.}~\bibnamefont {Tompkins}}, \bibinfo {author} {\bibfnamefont
  {J.}~\bibnamefont {Turner}}, \bibinfo {author} {\bibfnamefont
  {J.}~\bibnamefont {Welch}}, \bibinfo {author} {\bibfnamefont
  {W.}~\bibnamefont {White}}, \bibinfo {author} {\bibfnamefont
  {J.}~\bibnamefont {Wu}}, \bibinfo {author} {\bibfnamefont {G.}~\bibnamefont
  {Yocky}},\ and\ \bibinfo {author} {\bibfnamefont {G.}~\bibnamefont {J.}},\
  }\href@noop {} {\bibfield  {journal} {\bibinfo  {journal} {Nature Photonics}\
  }\textbf {\bibinfo {volume} {4}},\ \bibinfo {pages} {641} (\bibinfo {year}
  {2010})}\BibitemShut {NoStop}%
\bibitem [{\citenamefont {McNeil}\ and\ \citenamefont
  {Thompson}(2010)}]{McNeilThompson_XFEL_2010}%
  \BibitemOpen
  \bibfield  {author} {\bibinfo {author} {\bibfnamefont {B.~W.~J.}\
  \bibnamefont {McNeil}}\ and\ \bibinfo {author} {\bibfnamefont {N.~R.}\
  \bibnamefont {Thompson}},\ }\href@noop {} {\bibfield  {journal} {\bibinfo
  {journal} {Nature Photonics}\ }\textbf {\bibinfo {volume} {4}},\ \bibinfo
  {pages} {814} (\bibinfo {year} {2010})}\BibitemShut {NoStop}%
\bibitem [{\citenamefont {Ayvazyan}\ \emph {et~al.}(2002)\citenamefont
  {Ayvazyan} \emph {et~al.}}]{Ayvazyan_2002_EPJD}%
  \BibitemOpen
  \bibfield  {author} {\bibinfo {author} {\bibfnamefont {V.}~\bibnamefont
  {Ayvazyan}} \emph {et~al.},\ }\href@noop {} {\bibfield  {journal} {\bibinfo
  {journal} {Eur. Phys. J. D}\ }\textbf {\bibinfo {volume} {20}},\ \bibinfo
  {pages} {149} (\bibinfo {year} {2002})}\BibitemShut {NoStop}%
\bibitem [{\citenamefont {Yabashi}\ and\ \citenamefont
  {Tanaka}(2017)}]{Yabashi2017_NatPhotonics}%
  \BibitemOpen
  \bibfield  {author} {\bibinfo {author} {\bibfnamefont {M.}~\bibnamefont
  {Yabashi}}\ and\ \bibinfo {author} {\bibfnamefont {H.}~\bibnamefont
  {Tanaka}},\ }\href@noop {} {\bibfield  {journal} {\bibinfo  {journal} {Nature
  Photonics}\ }\textbf {\bibinfo {volume} {11}},\ \bibinfo {pages} {12}
  (\bibinfo {year} {2017})}\BibitemShut {NoStop}%
\bibitem [{\citenamefont {{Olive et al. (Particle Data
  Group)}}(2014)}]{ParticleDataGroup2014}%
  \BibitemOpen
  \bibfield  {author} {\bibinfo {author} {\bibfnamefont {K.~A.}\ \bibnamefont
  {{Olive et al. (Particle Data Group)}}},\ }\href@noop {} {\bibfield
  {journal} {\bibinfo  {journal} {Chin. Phys. C}\ }\textbf {\bibinfo {volume}
  {38}},\ \bibinfo {pages} {090001} (\bibinfo {year} {2014})}\BibitemShut
  {NoStop}%
\bibitem [{\citenamefont {Lindhard}(1965)}]{Lindhard}%
  \BibitemOpen
  \bibfield  {author} {\bibinfo {author} {\bibfnamefont {J.}~\bibnamefont
  {Lindhard}},\ }\href@noop {} {\bibfield  {journal} {\bibinfo  {journal} {K.
  Dan. Vidensk. Selsk. Mat. Fys. Medd.}\ }\textbf {\bibinfo {volume} {34}},\
  \bibinfo {pages} {1} (\bibinfo {year} {1965})}\BibitemShut {NoStop}%
\bibitem [{\citenamefont {Bandiera}\ \emph {et~al.}(2015)\citenamefont
  {Bandiera}, \citenamefont {Bagli}, \citenamefont {Germogli}, \citenamefont
  {Guidi}, \citenamefont {Mazzolari}, \citenamefont {Backe}, \citenamefont
  {Lauth}, \citenamefont {Berra}, \citenamefont {Lietti}, \citenamefont
  {Prest}, \citenamefont {De~Salvador}, \citenamefont {Vallazza},\ and\
  \citenamefont {Tikhomirov}}]{Backe_EtAl_PRL_115_025504_2015}%
  \BibitemOpen
  \bibfield  {author} {\bibinfo {author} {\bibfnamefont {L.}~\bibnamefont
  {Bandiera}}, \bibinfo {author} {\bibfnamefont {E.}~\bibnamefont {Bagli}},
  \bibinfo {author} {\bibfnamefont {G.}~\bibnamefont {Germogli}}, \bibinfo
  {author} {\bibfnamefont {V.}~\bibnamefont {Guidi}}, \bibinfo {author}
  {\bibfnamefont {M.}~\bibnamefont {Mazzolari}}, \bibinfo {author}
  {\bibfnamefont {H.}~\bibnamefont {Backe}}, \bibinfo {author} {\bibfnamefont
  {W.}~\bibnamefont {Lauth}}, \bibinfo {author} {\bibfnamefont
  {A.}~\bibnamefont {Berra}}, \bibinfo {author} {\bibfnamefont
  {D.}~\bibnamefont {Lietti}}, \bibinfo {author} {\bibfnamefont
  {M.}~\bibnamefont {Prest}}, \bibinfo {author} {\bibfnamefont
  {D.}~\bibnamefont {De~Salvador}}, \bibinfo {author} {\bibfnamefont
  {E.}~\bibnamefont {Vallazza}},\ and\ \bibinfo {author} {\bibfnamefont
  {V.}~\bibnamefont {Tikhomirov}},\ }\href@noop {} {\bibfield  {journal}
  {\bibinfo  {journal} {Phys. Rev. Lett.}\ }\textbf {\bibinfo {volume} {115}},\
  \bibinfo {pages} {025504} (\bibinfo {year} {2015})}\BibitemShut {NoStop}%
\bibitem [{\citenamefont {Kumakhov}(1976)}]{ChRad:Kumakhov1976}%
  \BibitemOpen
  \bibfield  {author} {\bibinfo {author} {\bibfnamefont {M.~A.}\ \bibnamefont
  {Kumakhov}},\ }\href@noop {} {\bibfield  {journal} {\bibinfo  {journal}
  {Phys. Lett.}\ }\textbf {\bibinfo {volume} {57A}},\ \bibinfo {pages} {17}
  (\bibinfo {year} {1976})}\BibitemShut {NoStop}%
\bibitem [{\citenamefont {Bezchastnov}\ \emph {et~al.}(2014)\citenamefont
  {Bezchastnov}, \citenamefont {Korol},\ and\ \citenamefont
  {Solov'yov}}]{BezchastnovKorolSolovyov2014}%
  \BibitemOpen
  \bibfield  {author} {\bibinfo {author} {\bibfnamefont {V.~G.}\ \bibnamefont
  {Bezchastnov}}, \bibinfo {author} {\bibfnamefont {A.~V.}\ \bibnamefont
  {Korol}},\ and\ \bibinfo {author} {\bibfnamefont {A.~V.}\ \bibnamefont
  {Solov'yov}},\ }\href@noop {} {\bibfield  {journal} {\bibinfo  {journal} {J.
  Phys. B: At. Mol. Opt. Phys.}\ }\textbf {\bibinfo {volume} {47}},\ \bibinfo
  {pages} {195401} (\bibinfo {year} {2014})}\BibitemShut {NoStop}%
\bibitem [{\citenamefont {Ginzburg}(1947)}]{Ginzburg_1947}%
  \BibitemOpen
  \bibfield  {author} {\bibinfo {author} {\bibfnamefont {V.~L.}\ \bibnamefont
  {Ginzburg}},\ }\href@noop {} {\bibfield  {journal} {\bibinfo  {journal} {Izv.
  Akad. Nauk SSSR (in Russian)}\ }\textbf {\bibinfo {volume} {11}},\ \bibinfo
  {pages} {165} (\bibinfo {year} {1947})}\BibitemShut {NoStop}%
\bibitem [{\citenamefont {Motz}(1951)}]{Motz_1951}%
  \BibitemOpen
  \bibfield  {author} {\bibinfo {author} {\bibfnamefont {H.}~\bibnamefont
  {Motz}},\ }\href@noop {} {\bibfield  {journal} {\bibinfo  {journal} {J. Appl.
  Phys.}\ }\textbf {\bibinfo {volume} {22}},\ \bibinfo {pages} {527} (\bibinfo
  {year} {1951})}\BibitemShut {NoStop}%
\bibitem [{\citenamefont {Rullhusen}\ \emph {et~al.}(1998)\citenamefont
  {Rullhusen}, \citenamefont {Artru},\ and\ \citenamefont
  {Dhez}}]{RullhusenArtruDhez}%
  \BibitemOpen
  \bibfield  {author} {\bibinfo {author} {\bibfnamefont {P.}~\bibnamefont
  {Rullhusen}}, \bibinfo {author} {\bibfnamefont {X.}~\bibnamefont {Artru}},\
  and\ \bibinfo {author} {\bibfnamefont {P.}~\bibnamefont {Dhez}},\ }\href@noop
  {} {\emph {\bibinfo {title} {Novel Radiation Sources using Relativistic
  Electrons}}}\ (\bibinfo  {publisher} {World Scientific, Singapore},\ \bibinfo
  {year} {1998})\BibitemShut {NoStop}%
\bibitem [{\citenamefont {Korol}\ \emph {et~al.}(2001)\citenamefont {Korol},
  \citenamefont {Solov'yov},\ and\ \citenamefont {Greiner}}]{Dechan01}%
  \BibitemOpen
  \bibfield  {author} {\bibinfo {author} {\bibfnamefont {A.~V.}\ \bibnamefont
  {Korol}}, \bibinfo {author} {\bibfnamefont {A.~V.}\ \bibnamefont
  {Solov'yov}},\ and\ \bibinfo {author} {\bibfnamefont {W.}~\bibnamefont
  {Greiner}},\ }\href@noop {} {\bibfield  {journal} {\bibinfo  {journal} {J.
  Phys. G: Nucl. Part. Phys.}\ }\textbf {\bibinfo {volume} {27}},\ \bibinfo
  {pages} {95} (\bibinfo {year} {2001})}\BibitemShut {NoStop}%
\bibitem [{\citenamefont {Korol}\ \emph {et~al.}(2000)\citenamefont {Korol},
  \citenamefont {Solov'yov},\ and\ \citenamefont {Greiner}}]{EnLoss00}%
  \BibitemOpen
  \bibfield  {author} {\bibinfo {author} {\bibfnamefont {A.~V.}\ \bibnamefont
  {Korol}}, \bibinfo {author} {\bibfnamefont {A.~V.}\ \bibnamefont
  {Solov'yov}},\ and\ \bibinfo {author} {\bibfnamefont {W.}~\bibnamefont
  {Greiner}},\ }\href@noop {} {\bibfield  {journal} {\bibinfo  {journal} {Int.
  J. Mod. Phys. E}\ }\textbf {\bibinfo {volume} {9}},\ \bibinfo {pages} {77}
  (\bibinfo {year} {2000})}\BibitemShut {NoStop}%
\bibitem [{\citenamefont {Baier}\ \emph {et~al.}(1998)\citenamefont {Baier},
  \citenamefont {Katkov},\ and\ \citenamefont {Strakhovenko}}]{Baier}%
  \BibitemOpen
  \bibfield  {author} {\bibinfo {author} {\bibfnamefont {V.~N.}\ \bibnamefont
  {Baier}}, \bibinfo {author} {\bibfnamefont {V.~M.}\ \bibnamefont {Katkov}},\
  and\ \bibinfo {author} {\bibfnamefont {V.~M.}\ \bibnamefont {Strakhovenko}},\
  }\href@noop {} {\emph {\bibinfo {title} {Electromagnetic Processes at High
  Energies in Oriented Single Crystals}}}\ (\bibinfo  {publisher} {World
  Scientific, Singapore},\ \bibinfo {year} {1998})\BibitemShut {NoStop}%
\bibitem [{\citenamefont {Sushko}\ \emph {et~al.}(2015)\citenamefont {Sushko},
  \citenamefont {Korol},\ and\ \citenamefont {Solov'yov}}]{Sushko_AK_AS_2015}%
  \BibitemOpen
  \bibfield  {author} {\bibinfo {author} {\bibfnamefont {G.~B.}\ \bibnamefont
  {Sushko}}, \bibinfo {author} {\bibfnamefont {A.~V.}\ \bibnamefont {Korol}},\
  and\ \bibinfo {author} {\bibfnamefont {A.~V.}\ \bibnamefont {Solov'yov}},\
  }\href@noop {} {\bibfield  {journal} {\bibinfo  {journal} {Nucl. Instrum.
  Meth. B}\ }\textbf {\bibinfo {volume} {355}},\ \bibinfo {pages} {39}
  (\bibinfo {year} {2015})}\BibitemShut {NoStop}%
\bibitem [{\citenamefont {Korol}\ \emph {et~al.}(2016)\citenamefont {Korol},
  \citenamefont {Bezchastnov}, \citenamefont {Sushko},\ and\ \citenamefont
  {Solov'yov}}]{KorolBezchastnovSushkoSolovyov2016}%
  \BibitemOpen
  \bibfield  {author} {\bibinfo {author} {\bibfnamefont {A.~V.}\ \bibnamefont
  {Korol}}, \bibinfo {author} {\bibfnamefont {V.~G.}\ \bibnamefont
  {Bezchastnov}}, \bibinfo {author} {\bibfnamefont {G.}~\bibnamefont
  {Sushko}},\ and\ \bibinfo {author} {\bibfnamefont {A.~V.}\ \bibnamefont
  {Solov'yov}},\ }\href@noop {} {\bibfield  {journal} {\bibinfo  {journal}
  {Nucl. Instrum. Meth. B}\ }\textbf {\bibinfo {volume} {387}},\ \bibinfo
  {pages} {41} (\bibinfo {year} {2016})}\BibitemShut {NoStop}%
\bibitem [{\citenamefont {Shen}\ \emph {et~al.}(2018)\citenamefont {Shen},
  \citenamefont {Zhao}, \citenamefont {Zhang}, \citenamefont {Sushko},
  \citenamefont {Korol},\ and\ \citenamefont
  {Solov'yov}}]{Shen_2018_NIMB.424.26}%
  \BibitemOpen
  \bibfield  {author} {\bibinfo {author} {\bibfnamefont {H.}~\bibnamefont
  {Shen}}, \bibinfo {author} {\bibfnamefont {Q.}~\bibnamefont {Zhao}}, \bibinfo
  {author} {\bibfnamefont {F.~S.}\ \bibnamefont {Zhang}}, \bibinfo {author}
  {\bibfnamefont {G.~B.}\ \bibnamefont {Sushko}}, \bibinfo {author}
  {\bibfnamefont {A.~V.}\ \bibnamefont {Korol}},\ and\ \bibinfo {author}
  {\bibfnamefont {A.~V.}\ \bibnamefont {Solov'yov}},\ }\href@noop {} {\bibfield
   {journal} {\bibinfo  {journal} {Nucl. Instrum. Meth. B}\ }\textbf {\bibinfo
  {volume} {424}},\ \bibinfo {pages} {26} (\bibinfo {year} {2018})}\BibitemShut
  {NoStop}%
\bibitem [{\citenamefont {Pavlov}\ \emph {et~al.}(2019)\citenamefont {Pavlov},
  \citenamefont {Korol}, \citenamefont {Ivanov},\ and\ \citenamefont
  {Solov'yov}}]{Pavlov_2019_JPB.52.11LT01}%
  \BibitemOpen
  \bibfield  {author} {\bibinfo {author} {\bibfnamefont {A.~V.}\ \bibnamefont
  {Pavlov}}, \bibinfo {author} {\bibfnamefont {A.~V.}\ \bibnamefont {Korol}},
  \bibinfo {author} {\bibfnamefont {V.~K.}\ \bibnamefont {Ivanov}},\ and\
  \bibinfo {author} {\bibfnamefont {A.~V.}\ \bibnamefont {Solov'yov}},\
  }\href@noop {} {\bibfield  {journal} {\bibinfo  {journal} {J. Phys. B: At.
  Mol. Opt. Phys.}\ }\textbf {\bibinfo {volume} {52}},\ \bibinfo {pages}
  {11LT01} (\bibinfo {year} {2019})}\BibitemShut {NoStop}%
\bibitem [{\citenamefont {Pavlov}\ \emph {et~al.}(2020)\citenamefont {Pavlov},
  \citenamefont {Korol}, \citenamefont {Ivanov},\ and\ \citenamefont
  {Solov'yov}}]{Pavlov_2020_EPJD.74.21}%
  \BibitemOpen
  \bibfield  {author} {\bibinfo {author} {\bibfnamefont {A.~V.}\ \bibnamefont
  {Pavlov}}, \bibinfo {author} {\bibfnamefont {A.~V.}\ \bibnamefont {Korol}},
  \bibinfo {author} {\bibfnamefont {V.~K.}\ \bibnamefont {Ivanov}},\ and\
  \bibinfo {author} {\bibfnamefont {A.~V.}\ \bibnamefont {Solov'yov}},\
  }\href@noop {} {\bibfield  {journal} {\bibinfo  {journal} {Eur. Phys. J. D}\
  }\textbf {\bibinfo {volume} {74}},\ \bibinfo {pages} {21} (\bibinfo {year}
  {2020})}\BibitemShut {NoStop}%
\bibitem [{\citenamefont {Cui}(2017)}]{Cui_Nanofabrication_book}%
  \BibitemOpen
  \bibfield  {author} {\bibinfo {author} {\bibfnamefont {Z.}~\bibnamefont
  {Cui}},\ }\href@noop {} {\emph {\bibinfo {title} {Nanofabrication.
  Principles, Capabilities and Limits}}}\ (\bibinfo  {publisher} {Springer
  International Publishing},\ \bibinfo {address} {Cham, Switzerland},\ \bibinfo
  {year} {2017})\BibitemShut {NoStop}%
\bibitem [{\citenamefont {Barth}\ \emph {et~al.}(2020)\citenamefont {Barth},
  \citenamefont {Huth},\ and\ \citenamefont
  {Jungwirth}}]{Barth2020_JMaterChemC}%
  \BibitemOpen
  \bibfield  {author} {\bibinfo {author} {\bibfnamefont {S.}~\bibnamefont
  {Barth}}, \bibinfo {author} {\bibfnamefont {M.}~\bibnamefont {Huth}},\ and\
  \bibinfo {author} {\bibfnamefont {F.}~\bibnamefont {Jungwirth}},\ }\href@noop
  {} {\bibfield  {journal} {\bibinfo  {journal} {J. Mater. Chem. C}\ }\textbf
  {\bibinfo {volume} {8}},\ \bibinfo {pages} {15884} (\bibinfo {year}
  {2020})}\BibitemShut {NoStop}%
\bibitem [{\citenamefont {Utke}\ \emph {et~al.}(2008)\citenamefont {Utke},
  \citenamefont {Hoffmann},\ and\ \citenamefont {Melngailis}}]{Utke2008}%
  \BibitemOpen
  \bibfield  {author} {\bibinfo {author} {\bibfnamefont {I.}~\bibnamefont
  {Utke}}, \bibinfo {author} {\bibfnamefont {P.}~\bibnamefont {Hoffmann}},\
  and\ \bibinfo {author} {\bibfnamefont {J.}~\bibnamefont {Melngailis}},\
  }\href@noop {} {\bibfield  {journal} {\bibinfo  {journal} {J. Vac. Sci.
  Technol. B}\ }\textbf {\bibinfo {volume} {26}},\ \bibinfo {pages} {1197}
  (\bibinfo {year} {2008})}\BibitemShut {NoStop}%
\bibitem [{\citenamefont {Huth}\ \emph {et~al.}(2018)\citenamefont {Huth},
  \citenamefont {Porrati},\ and\ \citenamefont {Dobrovolskiy}}]{Huth2018}%
  \BibitemOpen
  \bibfield  {author} {\bibinfo {author} {\bibfnamefont {M.}~\bibnamefont
  {Huth}}, \bibinfo {author} {\bibfnamefont {F.}~\bibnamefont {Porrati}},\ and\
  \bibinfo {author} {\bibfnamefont {O.~V.}\ \bibnamefont {Dobrovolskiy}},\
  }\href@noop {} {\bibfield  {journal} {\bibinfo  {journal} {Microelectron.
  Eng.}\ }\textbf {\bibinfo {volume} {185--186}},\ \bibinfo {pages} {9}
  (\bibinfo {year} {2018})}\BibitemShut {NoStop}%
\bibitem [{\citenamefont {Thorman}\ \emph {et~al.}(2015)\citenamefont
  {Thorman}, \citenamefont {{Ragesh Kumar}}, \citenamefont {Fairbrother},\ and\
  \citenamefont {Ing{\'{o}}lfsson}}]{Thorman2015}%
  \BibitemOpen
  \bibfield  {author} {\bibinfo {author} {\bibfnamefont {R.~M.}\ \bibnamefont
  {Thorman}}, \bibinfo {author} {\bibfnamefont {T.~P.}\ \bibnamefont {{Ragesh
  Kumar}}}, \bibinfo {author} {\bibfnamefont {D.~H.}\ \bibnamefont
  {Fairbrother}},\ and\ \bibinfo {author} {\bibfnamefont {O.}~\bibnamefont
  {Ing{\'{o}}lfsson}},\ }\href@noop {} {\bibfield  {journal} {\bibinfo
  {journal} {Beilstein J. Nanotechnol.}\ }\textbf {\bibinfo {volume} {6}},\
  \bibinfo {pages} {1904} (\bibinfo {year} {2015})}\BibitemShut {NoStop}%
\bibitem [{\citenamefont {de~Vera}\ \emph {et~al.}(2020)\citenamefont
  {de~Vera}, \citenamefont {Azzolini}, \citenamefont {Sushko}, \citenamefont
  {Abril}, \citenamefont {Garcia-Molina}, \citenamefont {Dapor}, \citenamefont
  {Solov'yov},\ and\ \citenamefont {Solov'yov}}]{DeVera2020}%
  \BibitemOpen
  \bibfield  {author} {\bibinfo {author} {\bibfnamefont {P.}~\bibnamefont
  {de~Vera}}, \bibinfo {author} {\bibfnamefont {M.}~\bibnamefont {Azzolini}},
  \bibinfo {author} {\bibfnamefont {G.}~\bibnamefont {Sushko}}, \bibinfo
  {author} {\bibfnamefont {I.}~\bibnamefont {Abril}}, \bibinfo {author}
  {\bibfnamefont {R.}~\bibnamefont {Garcia-Molina}}, \bibinfo {author}
  {\bibfnamefont {M.}~\bibnamefont {Dapor}}, \bibinfo {author} {\bibfnamefont
  {I.~A.}\ \bibnamefont {Solov'yov}},\ and\ \bibinfo {author} {\bibfnamefont
  {A.~V.}\ \bibnamefont {Solov'yov}},\ }\href@noop {} {\bibfield  {journal}
  {\bibinfo  {journal} {Sci. Rep.}\ }\textbf {\bibinfo {volume} {10}},\
  \bibinfo {pages} {20827} (\bibinfo {year} {2020})}\BibitemShut {NoStop}%
\bibitem [{\citenamefont {Fowlkes}\ and\ \citenamefont
  {Rack}(2010)}]{Fowlkes2010}%
  \BibitemOpen
  \bibfield  {author} {\bibinfo {author} {\bibfnamefont {J.~D.}\ \bibnamefont
  {Fowlkes}}\ and\ \bibinfo {author} {\bibfnamefont {P.~D.}\ \bibnamefont
  {Rack}},\ }\href@noop {} {\bibfield  {journal} {\bibinfo  {journal} {ACS
  Nano}\ }\textbf {\bibinfo {volume} {4}},\ \bibinfo {pages} {1619} (\bibinfo
  {year} {2010})}\BibitemShut {NoStop}%
\bibitem [{\citenamefont {Sanz-Hern{\'{a}}ndez}\ and\ \citenamefont
  {Fern{\'{a}}ndez-Pacheco}(2017)}]{Sanz-Hernandez2017}%
  \BibitemOpen
  \bibfield  {author} {\bibinfo {author} {\bibfnamefont {D.}~\bibnamefont
  {Sanz-Hern{\'{a}}ndez}}\ and\ \bibinfo {author} {\bibfnamefont
  {A.}~\bibnamefont {Fern{\'{a}}ndez-Pacheco}},\ }\href@noop {} {\bibfield
  {journal} {\bibinfo  {journal} {Beilstein J. Nanotechnol.}\ }\textbf
  {\bibinfo {volume} {8}},\ \bibinfo {pages} {2151} (\bibinfo {year}
  {2017})}\BibitemShut {NoStop}%
\bibitem [{\citenamefont {Muthukumar}\ \emph {et~al.}(2012)\citenamefont
  {Muthukumar}, \citenamefont {Jeschke}, \citenamefont {Valent{\'{i}}},
  \citenamefont {Begun}, \citenamefont {Schwenk}, \citenamefont {Porrati},\
  and\ \citenamefont {Huth}}]{Muthukumar2012}%
  \BibitemOpen
  \bibfield  {author} {\bibinfo {author} {\bibfnamefont {K.}~\bibnamefont
  {Muthukumar}}, \bibinfo {author} {\bibfnamefont {H.~O.}\ \bibnamefont
  {Jeschke}}, \bibinfo {author} {\bibfnamefont {R.}~\bibnamefont
  {Valent{\'{i}}}}, \bibinfo {author} {\bibfnamefont {E.}~\bibnamefont
  {Begun}}, \bibinfo {author} {\bibfnamefont {J.}~\bibnamefont {Schwenk}},
  \bibinfo {author} {\bibfnamefont {F.}~\bibnamefont {Porrati}},\ and\ \bibinfo
  {author} {\bibfnamefont {M.}~\bibnamefont {Huth}},\ }\href@noop {} {\bibfield
   {journal} {\bibinfo  {journal} {Beilstein J. Nanotechnol.}\ }\textbf
  {\bibinfo {volume} {3}},\ \bibinfo {pages} {546} (\bibinfo {year}
  {2012})}\BibitemShut {NoStop}%
\bibitem [{\citenamefont {Muthukumar}\ \emph {et~al.}(2018)\citenamefont
  {Muthukumar}, \citenamefont {Jeschke},\ and\ \citenamefont
  {Valent{\'{i}}}}]{Muthukumar2018}%
  \BibitemOpen
  \bibfield  {author} {\bibinfo {author} {\bibfnamefont {K.}~\bibnamefont
  {Muthukumar}}, \bibinfo {author} {\bibfnamefont {H.~O.}\ \bibnamefont
  {Jeschke}},\ and\ \bibinfo {author} {\bibfnamefont {R.}~\bibnamefont
  {Valent{\'{i}}}},\ }\href@noop {} {\bibfield  {journal} {\bibinfo  {journal}
  {Beilstein J. Nanotechnol.}\ }\textbf {\bibinfo {volume} {9}},\ \bibinfo
  {pages} {711} (\bibinfo {year} {2018})}\BibitemShut {NoStop}%
\bibitem [{\citenamefont {Sushko}\ \emph
  {et~al.}(2016{\natexlab{a}})\citenamefont {Sushko}, \citenamefont
  {Solov'yov}, \citenamefont {Verkhovtsev}, \citenamefont {Volkov},\ and\
  \citenamefont {Solov'yov}}]{Sushko_2016_EPJD.70.12}%
  \BibitemOpen
  \bibfield  {author} {\bibinfo {author} {\bibfnamefont {G.~B.}\ \bibnamefont
  {Sushko}}, \bibinfo {author} {\bibfnamefont {I.~A.}\ \bibnamefont
  {Solov'yov}}, \bibinfo {author} {\bibfnamefont {A.~V.}\ \bibnamefont
  {Verkhovtsev}}, \bibinfo {author} {\bibfnamefont {S.~N.}\ \bibnamefont
  {Volkov}},\ and\ \bibinfo {author} {\bibfnamefont {A.~V.}\ \bibnamefont
  {Solov'yov}},\ }\href@noop {} {\bibfield  {journal} {\bibinfo  {journal}
  {Eur. Phys. J. D}\ }\textbf {\bibinfo {volume} {70}},\ \bibinfo {pages} {12}
  (\bibinfo {year} {2016}{\natexlab{a}})}\BibitemShut {NoStop}%
\bibitem [{\citenamefont {de~Vera}\ \emph {et~al.}(2019)\citenamefont
  {de~Vera}, \citenamefont {Verkhovtsev}, \citenamefont {Sushko},\ and\
  \citenamefont {Solov'yov}}]{deVera_2019_EPJD.73.215}%
  \BibitemOpen
  \bibfield  {author} {\bibinfo {author} {\bibfnamefont {P.}~\bibnamefont
  {de~Vera}}, \bibinfo {author} {\bibfnamefont {A.}~\bibnamefont
  {Verkhovtsev}}, \bibinfo {author} {\bibfnamefont {G.}~\bibnamefont
  {Sushko}},\ and\ \bibinfo {author} {\bibfnamefont {A.~V.}\ \bibnamefont
  {Solov'yov}},\ }\href@noop {} {\bibfield  {journal} {\bibinfo  {journal}
  {Eur. Phys. J. D}\ }\textbf {\bibinfo {volume} {73}},\ \bibinfo {pages} {215}
  (\bibinfo {year} {2019})}\BibitemShut {NoStop}%
\bibitem [{\citenamefont {Sushko}\ \emph
  {et~al.}(2016{\natexlab{b}})\citenamefont {Sushko}, \citenamefont
  {Solov'yov},\ and\ \citenamefont {Solov'yov}}]{Sushko_IS_AS_FEBID_2016}%
  \BibitemOpen
  \bibfield  {author} {\bibinfo {author} {\bibfnamefont {G.~B.}\ \bibnamefont
  {Sushko}}, \bibinfo {author} {\bibfnamefont {I.~A.}\ \bibnamefont
  {Solov'yov}},\ and\ \bibinfo {author} {\bibfnamefont {A.~V.}\ \bibnamefont
  {Solov'yov}},\ }\href@noop {} {\bibfield  {journal} {\bibinfo  {journal}
  {Eur. Phys. J. D}\ }\textbf {\bibinfo {volume} {70}},\ \bibinfo {pages} {217}
  (\bibinfo {year} {2016}{\natexlab{b}})}\BibitemShut {NoStop}%
\bibitem [{\citenamefont {Friis}\ \emph {et~al.}(2020)\citenamefont {Friis},
  \citenamefont {Verkhovtsev}, \citenamefont {Solov'yov},\ and\ \citenamefont
  {Solov'yov}}]{Friis2020}%
  \BibitemOpen
  \bibfield  {author} {\bibinfo {author} {\bibfnamefont {I.}~\bibnamefont
  {Friis}}, \bibinfo {author} {\bibfnamefont {A.}~\bibnamefont {Verkhovtsev}},
  \bibinfo {author} {\bibfnamefont {I.~A.}\ \bibnamefont {Solov'yov}},\ and\
  \bibinfo {author} {\bibfnamefont {A.~V.}\ \bibnamefont {Solov'yov}},\
  }\href@noop {} {\bibfield  {journal} {\bibinfo  {journal} {J. Comput. Chem.}\
  }\textbf {\bibinfo {volume} {41}},\ \bibinfo {pages} {2429} (\bibinfo {year}
  {2020})}\BibitemShut {NoStop}%
\bibitem [{\citenamefont {Dapor}(2020)}]{Dapor_2020_MC}%
  \BibitemOpen
  \bibfield  {author} {\bibinfo {author} {\bibfnamefont {M.}~\bibnamefont
  {Dapor}},\ }\href@noop {} {\emph {\bibinfo {title} {{Transport of Energetic
  Electrons in Solids (3rd ed.)}}}}\ (\bibinfo  {publisher} {Springer
  International Publishing},\ \bibinfo {address} {Cham, Switzerland},\ \bibinfo
  {year} {2020})\BibitemShut {NoStop}%
\bibitem [{\citenamefont {Azzolini}\ \emph {et~al.}(2019)\citenamefont
  {Azzolini}, \citenamefont {Angelucci}, \citenamefont {Cimino}, \citenamefont
  {Larciprete}, \citenamefont {Pugno}, \citenamefont {Taioli},\ and\
  \citenamefont {Dapor}}]{Azzolini_2019_SEED}%
  \BibitemOpen
  \bibfield  {author} {\bibinfo {author} {\bibfnamefont {M.}~\bibnamefont
  {Azzolini}}, \bibinfo {author} {\bibfnamefont {M.}~\bibnamefont {Angelucci}},
  \bibinfo {author} {\bibfnamefont {R.}~\bibnamefont {Cimino}}, \bibinfo
  {author} {\bibfnamefont {R.}~\bibnamefont {Larciprete}}, \bibinfo {author}
  {\bibfnamefont {N.~M.}\ \bibnamefont {Pugno}}, \bibinfo {author}
  {\bibfnamefont {S.}~\bibnamefont {Taioli}},\ and\ \bibinfo {author}
  {\bibfnamefont {M.}~\bibnamefont {Dapor}},\ }\href@noop {} {\bibfield
  {journal} {\bibinfo  {journal} {J. Phys.: Condens. Matter}\ }\textbf
  {\bibinfo {volume} {31}},\ \bibinfo {pages} {055901} (\bibinfo {year}
  {2019})}\BibitemShut {NoStop}%
\bibitem [{\citenamefont {Porrati}\ \emph {et~al.}(2009)\citenamefont
  {Porrati}, \citenamefont {Sachser},\ and\ \citenamefont
  {Huth}}]{Porrati_2009_Nanotechnology}%
  \BibitemOpen
  \bibfield  {author} {\bibinfo {author} {\bibfnamefont {F.}~\bibnamefont
  {Porrati}}, \bibinfo {author} {\bibfnamefont {R.}~\bibnamefont {Sachser}},\
  and\ \bibinfo {author} {\bibfnamefont {M.}~\bibnamefont {Huth}},\ }\href@noop
  {} {\bibfield  {journal} {\bibinfo  {journal} {Nanotechnology}\ }\textbf
  {\bibinfo {volume} {20}},\ \bibinfo {pages} {195301} (\bibinfo {year}
  {2009})}\BibitemShut {NoStop}%
\bibitem [{\citenamefont {Linz}(2012)}]{Linz2012_IonBeams}%
  \BibitemOpen
  \bibinfo {editor} {\bibfnamefont {U.}~\bibnamefont {Linz}},\ ed.,\ \href@noop
  {} {\emph {\bibinfo {title} {Ion Beam Therapy: Fundamentals, Technology,
  Clinical Applications}}}\ (\bibinfo  {publisher} {Springer},\ \bibinfo {year}
  {2012})\BibitemShut {NoStop}%
\bibitem [{\citenamefont {Surdutovich}\ and\ \citenamefont
  {Solov'yov}(2019)}]{surdutovich2019multiscale}%
  \BibitemOpen
  \bibfield  {author} {\bibinfo {author} {\bibfnamefont {E.}~\bibnamefont
  {Surdutovich}}\ and\ \bibinfo {author} {\bibfnamefont {A.~V.}\ \bibnamefont
  {Solov'yov}},\ }\href@noop {} {\bibfield  {journal} {\bibinfo  {journal}
  {Cancer Nanotechnol.}\ }\textbf {\bibinfo {volume} {10}},\ \bibinfo {pages}
  {6} (\bibinfo {year} {2019})}\BibitemShut {NoStop}%
\bibitem [{\citenamefont {Alpen}(1998)}]{Alpen}%
  \BibitemOpen
  \bibfield  {author} {\bibinfo {author} {\bibfnamefont {E.~L.}\ \bibnamefont
  {Alpen}},\ }\href@noop {} {\emph {\bibinfo {title} {Radiation Biophysics}}}\
  (\bibinfo  {publisher} {Academic Press, San Diego, London, Boston, New York,
  Sydney, Tokyo, Toronto},\ \bibinfo {year} {1998})\BibitemShut {NoStop}%
\bibitem [{\citenamefont {Verkhovtsev}\ \emph {et~al.}(2016)\citenamefont
  {Verkhovtsev}, \citenamefont {Surdutovich},\ and\ \citenamefont
  {Solov'yov}}]{verkhovtsev2016multiscale}%
  \BibitemOpen
  \bibfield  {author} {\bibinfo {author} {\bibfnamefont {A.}~\bibnamefont
  {Verkhovtsev}}, \bibinfo {author} {\bibfnamefont {E.}~\bibnamefont
  {Surdutovich}},\ and\ \bibinfo {author} {\bibfnamefont {A.~V.}\ \bibnamefont
  {Solov'yov}},\ }\href@noop {} {\bibfield  {journal} {\bibinfo  {journal}
  {Sci. Rep.}\ }\textbf {\bibinfo {volume} {6}},\ \bibinfo {pages} {27654}
  (\bibinfo {year} {2016})}\BibitemShut {NoStop}%
\bibitem [{\citenamefont {Verkhovtsev}\ \emph {et~al.}(2019)\citenamefont
  {Verkhovtsev}, \citenamefont {Surdutovich},\ and\ \citenamefont
  {Solov'yov}}]{verkhovtsev2019phenomenon}%
  \BibitemOpen
  \bibfield  {author} {\bibinfo {author} {\bibfnamefont {A.}~\bibnamefont
  {Verkhovtsev}}, \bibinfo {author} {\bibfnamefont {E.}~\bibnamefont
  {Surdutovich}},\ and\ \bibinfo {author} {\bibfnamefont {A.~V.}\ \bibnamefont
  {Solov'yov}},\ }\href@noop {} {\bibfield  {journal} {\bibinfo  {journal}
  {Cancer Nanotechnol.}\ }\textbf {\bibinfo {volume} {10}},\ \bibinfo {pages}
  {4} (\bibinfo {year} {2019})}\BibitemShut {NoStop}%
\bibitem [{\citenamefont {Meesungnoen}\ \emph {et~al.}(2002)\citenamefont
  {Meesungnoen}, \citenamefont {Jay-Gerin}, \citenamefont {Filali-Mouhim},\
  and\ \citenamefont {Mankhetkorn}}]{Meesungnoen02}%
  \BibitemOpen
  \bibfield  {author} {\bibinfo {author} {\bibfnamefont {J.}~\bibnamefont
  {Meesungnoen}}, \bibinfo {author} {\bibfnamefont {J.-P.}\ \bibnamefont
  {Jay-Gerin}}, \bibinfo {author} {\bibfnamefont {A.}~\bibnamefont
  {Filali-Mouhim}},\ and\ \bibinfo {author} {\bibfnamefont {S.}~\bibnamefont
  {Mankhetkorn}},\ }\href@noop {} {\bibfield  {journal} {\bibinfo  {journal}
  {Radiat. Res.}\ }\textbf {\bibinfo {volume} {158}},\ \bibinfo {pages} {657}
  (\bibinfo {year} {2002})}\BibitemShut {NoStop}%
\bibitem [{\citenamefont {Nikjoo}\ \emph {et~al.}(2006)\citenamefont {Nikjoo},
  \citenamefont {Uehara}, \citenamefont {Emfietzoglou},\ and\ \citenamefont
  {Cucinotta}}]{Nikjoo06}%
  \BibitemOpen
  \bibfield  {author} {\bibinfo {author} {\bibfnamefont {H.}~\bibnamefont
  {Nikjoo}}, \bibinfo {author} {\bibfnamefont {S.}~\bibnamefont {Uehara}},
  \bibinfo {author} {\bibfnamefont {D.}~\bibnamefont {Emfietzoglou}},\ and\
  \bibinfo {author} {\bibfnamefont {F.~A.}\ \bibnamefont {Cucinotta}},\
  }\href@noop {} {\bibfield  {journal} {\bibinfo  {journal} {Radiat. Meas.}\
  }\textbf {\bibinfo {volume} {41}},\ \bibinfo {pages} {1052} (\bibinfo {year}
  {2006})}\BibitemShut {NoStop}%
\bibitem [{\citenamefont {Surdutovich}\ \emph {et~al.}(2011)\citenamefont
  {Surdutovich}, \citenamefont {Gallagher},\ and\ \citenamefont
  {Solov'yov}}]{precomplex}%
  \BibitemOpen
  \bibfield  {author} {\bibinfo {author} {\bibfnamefont {E.}~\bibnamefont
  {Surdutovich}}, \bibinfo {author} {\bibfnamefont {D.~C.}\ \bibnamefont
  {Gallagher}},\ and\ \bibinfo {author} {\bibfnamefont {A.~V.}\ \bibnamefont
  {Solov'yov}},\ }\href@noop {} {\bibfield  {journal} {\bibinfo  {journal}
  {Phys. Rev. E}\ }\textbf {\bibinfo {volume} {84}},\ \bibinfo {pages} {051918}
  (\bibinfo {year} {2011})}\BibitemShut {NoStop}%
\bibitem [{\citenamefont {Surdutovich}\ and\ \citenamefont
  {Solov'yov}(2012)}]{SurdutovichSolovyov_EPJD_2012}%
  \BibitemOpen
  \bibfield  {author} {\bibinfo {author} {\bibfnamefont {E.}~\bibnamefont
  {Surdutovich}}\ and\ \bibinfo {author} {\bibfnamefont {A.~V.}\ \bibnamefont
  {Solov'yov}},\ }\href@noop {} {\bibfield  {journal} {\bibinfo  {journal}
  {Eur. Phys. J. D}\ }\textbf {\bibinfo {volume} {66}},\ \bibinfo {pages} {206}
  (\bibinfo {year} {2012})}\BibitemShut {NoStop}%
\bibitem [{\citenamefont {Solov'yov}\ \emph {et~al.}(2009)\citenamefont
  {Solov'yov}, \citenamefont {Surdutovich}, \citenamefont {Scifoni},
  \citenamefont {Mishustin},\ and\ \citenamefont
  {Greiner}}]{Solovyov2009_IBCT}%
  \BibitemOpen
  \bibfield  {author} {\bibinfo {author} {\bibfnamefont {A.~V.}\ \bibnamefont
  {Solov'yov}}, \bibinfo {author} {\bibfnamefont {E.}~\bibnamefont
  {Surdutovich}}, \bibinfo {author} {\bibfnamefont {E.}~\bibnamefont
  {Scifoni}}, \bibinfo {author} {\bibfnamefont {I.}~\bibnamefont {Mishustin}},\
  and\ \bibinfo {author} {\bibfnamefont {W.}~\bibnamefont {Greiner}},\
  }\href@noop {} {\bibfield  {journal} {\bibinfo  {journal} {Phys. Rev. E}\
  }\textbf {\bibinfo {volume} {79}},\ \bibinfo {pages} {011909} (\bibinfo
  {year} {2009})}\BibitemShut {NoStop}%
\bibitem [{\citenamefont {Bug}\ \emph {et~al.}(2012)\citenamefont {Bug},
  \citenamefont {Surdutovich}, \citenamefont {Rabus}, \citenamefont
  {Rosenfeld},\ and\ \citenamefont {Solov'yov}}]{epjdmarion}%
  \BibitemOpen
  \bibfield  {author} {\bibinfo {author} {\bibfnamefont {M.}~\bibnamefont
  {Bug}}, \bibinfo {author} {\bibfnamefont {E.}~\bibnamefont {Surdutovich}},
  \bibinfo {author} {\bibfnamefont {H.}~\bibnamefont {Rabus}}, \bibinfo
  {author} {\bibfnamefont {A.~B.}\ \bibnamefont {Rosenfeld}},\ and\ \bibinfo
  {author} {\bibfnamefont {A.~V.}\ \bibnamefont {Solov'yov}},\ }\href@noop {}
  {\bibfield  {journal} {\bibinfo  {journal} {Eur. Phys. J. D}\ }\textbf
  {\bibinfo {volume} {66}},\ \bibinfo {pages} {291} (\bibinfo {year}
  {2012})}\BibitemShut {NoStop}%
\bibitem [{\citenamefont {Surdutovich}\ \emph {et~al.}(2009)\citenamefont
  {Surdutovich}, \citenamefont {Obolensky}, \citenamefont {Scifoni},
  \citenamefont {Pshenichnov}, \citenamefont {Mishustin}, \citenamefont
  {Solov'yov},\ and\ \citenamefont {Greiner}}]{epjd}%
  \BibitemOpen
  \bibfield  {author} {\bibinfo {author} {\bibfnamefont {E.}~\bibnamefont
  {Surdutovich}}, \bibinfo {author} {\bibfnamefont {O.}~\bibnamefont
  {Obolensky}}, \bibinfo {author} {\bibfnamefont {E.}~\bibnamefont {Scifoni}},
  \bibinfo {author} {\bibfnamefont {I.}~\bibnamefont {Pshenichnov}}, \bibinfo
  {author} {\bibfnamefont {I.}~\bibnamefont {Mishustin}}, \bibinfo {author}
  {\bibfnamefont {A.}~\bibnamefont {Solov'yov}},\ and\ \bibinfo {author}
  {\bibfnamefont {W.}~\bibnamefont {Greiner}},\ }\href@noop {} {\bibfield
  {journal} {\bibinfo  {journal} {Eur. Phys. J. D}\ }\textbf {\bibinfo {volume}
  {51}},\ \bibinfo {pages} {63} (\bibinfo {year} {2009})}\BibitemShut {NoStop}%
\bibitem [{\citenamefont {Surdutovich}\ \emph {et~al.}(2013)\citenamefont
  {Surdutovich}, \citenamefont {Yakubovich},\ and\ \citenamefont
  {Solov'yov}}]{surdutovich2013biodamage}%
  \BibitemOpen
  \bibfield  {author} {\bibinfo {author} {\bibfnamefont {E.}~\bibnamefont
  {Surdutovich}}, \bibinfo {author} {\bibfnamefont {A.~V.}\ \bibnamefont
  {Yakubovich}},\ and\ \bibinfo {author} {\bibfnamefont {A.~V.}\ \bibnamefont
  {Solov'yov}},\ }\href@noop {} {\bibfield  {journal} {\bibinfo  {journal}
  {Sci. Rep.}\ }\textbf {\bibinfo {volume} {3}},\ \bibinfo {pages} {1289}
  (\bibinfo {year} {2013})}\BibitemShut {NoStop}%
\bibitem [{\citenamefont {Park}\ \emph {et~al.}(2011)\citenamefont {Park},
  \citenamefont {Li}, \citenamefont {Cloutier}, \citenamefont {Sanche},\ and\
  \citenamefont {Wagner}}]{Sanche11}%
  \BibitemOpen
  \bibfield  {author} {\bibinfo {author} {\bibfnamefont {Y.}~\bibnamefont
  {Park}}, \bibinfo {author} {\bibfnamefont {Z.}~\bibnamefont {Li}}, \bibinfo
  {author} {\bibfnamefont {P.}~\bibnamefont {Cloutier}}, \bibinfo {author}
  {\bibfnamefont {L.}~\bibnamefont {Sanche}},\ and\ \bibinfo {author}
  {\bibfnamefont {J.}~\bibnamefont {Wagner}},\ }\href@noop {} {\bibfield
  {journal} {\bibinfo  {journal} {Radiat. Res.}\ }\textbf {\bibinfo {volume}
  {175}},\ \bibinfo {pages} {240} (\bibinfo {year} {2011})}\BibitemShut
  {NoStop}%
\bibitem [{\citenamefont {von Sonntag}(1987)}]{hyd2}%
  \BibitemOpen
  \bibfield  {author} {\bibinfo {author} {\bibfnamefont {C.}~\bibnamefont {von
  Sonntag}},\ }\href@noop {} {\emph {\bibinfo {title} {The Chemical Basis of
  Radiation Biology}}}\ (\bibinfo  {publisher} {Taylor \& Francis, London},\
  \bibinfo {year} {1987})\BibitemShut {NoStop}%
\bibitem [{\citenamefont {Smyth}\ and\ \citenamefont
  {Kohanoff}(2012)}]{Kohanoff2012}%
  \BibitemOpen
  \bibfield  {author} {\bibinfo {author} {\bibfnamefont {M.}~\bibnamefont
  {Smyth}}\ and\ \bibinfo {author} {\bibfnamefont {J.}~\bibnamefont
  {Kohanoff}},\ }\href@noop {} {\bibfield  {journal} {\bibinfo  {journal} {J.
  Am. Chem. Soc.}\ }\textbf {\bibinfo {volume} {134}},\ \bibinfo {pages} {9122}
  (\bibinfo {year} {2012})}\BibitemShut {NoStop}%
\bibitem [{\citenamefont {Toulemonde}\ \emph {et~al.}(2009)\citenamefont
  {Toulemonde}, \citenamefont {Surdutovich},\ and\ \citenamefont
  {Solov'yov}}]{Toulemonde2009_PRE}%
  \BibitemOpen
  \bibfield  {author} {\bibinfo {author} {\bibfnamefont {M.}~\bibnamefont
  {Toulemonde}}, \bibinfo {author} {\bibfnamefont {E.}~\bibnamefont
  {Surdutovich}},\ and\ \bibinfo {author} {\bibfnamefont {A.~V.}\ \bibnamefont
  {Solov'yov}},\ }\href@noop {} {\bibfield  {journal} {\bibinfo  {journal}
  {Phys. Rev. E}\ }\textbf {\bibinfo {volume} {80}},\ \bibinfo {pages} {031913}
  (\bibinfo {year} {2009})}\BibitemShut {NoStop}%
\bibitem [{\citenamefont {Surdutovich}\ and\ \citenamefont
  {Solov'yov}(2010)}]{surdutovich2010shock}%
  \BibitemOpen
  \bibfield  {author} {\bibinfo {author} {\bibfnamefont {E.}~\bibnamefont
  {Surdutovich}}\ and\ \bibinfo {author} {\bibfnamefont {A.~V.}\ \bibnamefont
  {Solov'yov}},\ }\href@noop {} {\bibfield  {journal} {\bibinfo  {journal}
  {Phys. Rev. E}\ }\textbf {\bibinfo {volume} {82}},\ \bibinfo {pages} {051915}
  (\bibinfo {year} {2010})}\BibitemShut {NoStop}%
\bibitem [{\citenamefont {Yakubovich}\ \emph {et~al.}(2011)\citenamefont
  {Yakubovich}, \citenamefont {Surdutovich},\ and\ \citenamefont
  {Solov'yov}}]{Yakubovich_2011_AIP.1344.230}%
  \BibitemOpen
  \bibfield  {author} {\bibinfo {author} {\bibfnamefont {A.~V.}\ \bibnamefont
  {Yakubovich}}, \bibinfo {author} {\bibfnamefont {E.}~\bibnamefont
  {Surdutovich}},\ and\ \bibinfo {author} {\bibfnamefont {A.~V.}\ \bibnamefont
  {Solov'yov}},\ }\href@noop {} {\bibfield  {journal} {\bibinfo  {journal} {AIP
  Conf. Proc.}\ }\textbf {\bibinfo {volume} {1344}},\ \bibinfo {pages} {230}
  (\bibinfo {year} {2011})}\BibitemShut {NoStop}%
\bibitem [{\citenamefont {de~Vera}\ \emph {et~al.}(2016)\citenamefont
  {de~Vera}, \citenamefont {Mason}, \citenamefont {Currell},\ and\
  \citenamefont {Solov'yov}}]{devera2016molecular}%
  \BibitemOpen
  \bibfield  {author} {\bibinfo {author} {\bibfnamefont {P.}~\bibnamefont
  {de~Vera}}, \bibinfo {author} {\bibfnamefont {N.~J.}\ \bibnamefont {Mason}},
  \bibinfo {author} {\bibfnamefont {F.~J.}\ \bibnamefont {Currell}},\ and\
  \bibinfo {author} {\bibfnamefont {A.~V.}\ \bibnamefont {Solov'yov}},\
  }\href@noop {} {\bibfield  {journal} {\bibinfo  {journal} {Eur. Phys. J. D}\
  }\textbf {\bibinfo {volume} {70}},\ \bibinfo {pages} {183} (\bibinfo {year}
  {2016})}\BibitemShut {NoStop}%
\bibitem [{\citenamefont {Fraile}\ \emph {et~al.}(2019)\citenamefont {Fraile},
  \citenamefont {Smyth}, \citenamefont {Kohanoff},\ and\ \citenamefont
  {Solov'yov}}]{fraile2019first}%
  \BibitemOpen
  \bibfield  {author} {\bibinfo {author} {\bibfnamefont {A.}~\bibnamefont
  {Fraile}}, \bibinfo {author} {\bibfnamefont {M.}~\bibnamefont {Smyth}},
  \bibinfo {author} {\bibfnamefont {J.}~\bibnamefont {Kohanoff}},\ and\
  \bibinfo {author} {\bibfnamefont {A.~V.}\ \bibnamefont {Solov'yov}},\
  }\href@noop {} {\bibfield  {journal} {\bibinfo  {journal} {J. Chem. Phys.}\
  }\textbf {\bibinfo {volume} {150}},\ \bibinfo {pages} {015101} (\bibinfo
  {year} {2019})}\BibitemShut {NoStop}%
\bibitem [{\citenamefont {Bottl{\"a}nder}\ \emph {et~al.}(2015)\citenamefont
  {Bottl{\"a}nder}, \citenamefont {M{\"u}cksch},\ and\ \citenamefont
  {Urbassek}}]{bottlander2015effect}%
  \BibitemOpen
  \bibfield  {author} {\bibinfo {author} {\bibfnamefont {D.}~\bibnamefont
  {Bottl{\"a}nder}}, \bibinfo {author} {\bibfnamefont {C.}~\bibnamefont
  {M{\"u}cksch}},\ and\ \bibinfo {author} {\bibfnamefont {H.~M.}\ \bibnamefont
  {Urbassek}},\ }\href@noop {} {\bibfield  {journal} {\bibinfo  {journal}
  {Nucl. Instrum. Meth. B}\ }\textbf {\bibinfo {volume} {365}},\ \bibinfo
  {pages} {622} (\bibinfo {year} {2015})}\BibitemShut {NoStop}%
\bibitem [{\citenamefont {Friis}\ \emph {et~al.}(2021)\citenamefont {Friis},
  \citenamefont {Verkhovtsev}, \citenamefont {Solov'yov},\ and\ \citenamefont
  {Solov'yov}}]{Friis2021_SWdamage}%
  \BibitemOpen
  \bibfield  {author} {\bibinfo {author} {\bibfnamefont {I.}~\bibnamefont
  {Friis}}, \bibinfo {author} {\bibfnamefont {A.~V.}\ \bibnamefont
  {Verkhovtsev}}, \bibinfo {author} {\bibfnamefont {I.~A.}\ \bibnamefont
  {Solov'yov}},\ and\ \bibinfo {author} {\bibfnamefont {A.~V.}\ \bibnamefont
  {Solov'yov}},\ }\href@noop {} {\bibfield  {journal} {\bibinfo  {journal}
  {arXiv:2103.10187 [physics.bio-ph]}\ } (\bibinfo {year} {2021})}\BibitemShut
  {NoStop}%
\bibitem [{\citenamefont {Surdutovich}\ and\ \citenamefont
  {Solov'yov}(2015)}]{Surdutovich_2015_EPJD.69.193}%
  \BibitemOpen
  \bibfield  {author} {\bibinfo {author} {\bibfnamefont {E.}~\bibnamefont
  {Surdutovich}}\ and\ \bibinfo {author} {\bibfnamefont {A.~V.}\ \bibnamefont
  {Solov'yov}},\ }\href@noop {} {\bibfield  {journal} {\bibinfo  {journal}
  {Eur. Phys. J. D}\ }\textbf {\bibinfo {volume} {69}},\ \bibinfo {pages} {193}
  (\bibinfo {year} {2015})}\BibitemShut {NoStop}%
\bibitem [{\citenamefont {de~Vera}\ \emph {et~al.}(2018)\citenamefont
  {de~Vera}, \citenamefont {Surdutovich}, \citenamefont {Mason}, \citenamefont
  {Currell},\ and\ \citenamefont {Solov'yov}}]{deVera_2018_EPJD.72.147}%
  \BibitemOpen
  \bibfield  {author} {\bibinfo {author} {\bibfnamefont {P.}~\bibnamefont
  {de~Vera}}, \bibinfo {author} {\bibfnamefont {E.}~\bibnamefont
  {Surdutovich}}, \bibinfo {author} {\bibfnamefont {N.~J.}\ \bibnamefont
  {Mason}}, \bibinfo {author} {\bibfnamefont {F.~J.}\ \bibnamefont {Currell}},\
  and\ \bibinfo {author} {\bibfnamefont {A.~V.}\ \bibnamefont {Solov'yov}},\
  }\href@noop {} {\bibfield  {journal} {\bibinfo  {journal} {Eur. Phys. J. D}\
  }\textbf {\bibinfo {volume} {72}},\ \bibinfo {pages} {147} (\bibinfo {year}
  {2018})}\BibitemShut {NoStop}%
\bibitem [{\citenamefont {Durante}\ and\ \citenamefont
  {Cucinotta}(2011)}]{durante2011physical}%
  \BibitemOpen
  \bibfield  {author} {\bibinfo {author} {\bibfnamefont {M.}~\bibnamefont
  {Durante}}\ and\ \bibinfo {author} {\bibfnamefont {F.~A.}\ \bibnamefont
  {Cucinotta}},\ }\href@noop {} {\bibfield  {journal} {\bibinfo  {journal}
  {Rev. Mod. Phys.}\ }\textbf {\bibinfo {volume} {83}},\ \bibinfo {pages}
  {1245} (\bibinfo {year} {2011})}\BibitemShut {NoStop}%
\bibitem [{\citenamefont {Kronenberg}\ and\ \citenamefont
  {Cucinotta}(2012)}]{Kronenberg2012_HealthPhys.103.556}%
  \BibitemOpen
  \bibfield  {author} {\bibinfo {author} {\bibfnamefont {A.}~\bibnamefont
  {Kronenberg}}\ and\ \bibinfo {author} {\bibfnamefont {F.~A.}\ \bibnamefont
  {Cucinotta}},\ }\href@noop {} {\bibfield  {journal} {\bibinfo  {journal}
  {Health Phys.}\ }\textbf {\bibinfo {volume} {103}},\ \bibinfo {pages} {556}
  (\bibinfo {year} {2012})}\BibitemShut {NoStop}%
\bibitem [{\citenamefont {MacKerell}\ \emph {et~al.}(1998)\citenamefont
  {MacKerell}, \citenamefont {Jr.}, \citenamefont {Bashford}, \citenamefont
  {Bellott}, \citenamefont {Dunbrack}, \citenamefont {Jr.}, \citenamefont
  {Evanseck}, \citenamefont {Field}, \citenamefont {Fischer}, \citenamefont
  {Gao}, \citenamefont {Guo}, \citenamefont {Ha}, \citenamefont
  {Joseph-McCarthy}, \citenamefont {Kuchnir}, \citenamefont {Kuczera},
  \citenamefont {Lau}, \citenamefont {Mattos}, \citenamefont {Michnick},
  \citenamefont {Ngo}, \citenamefont {Nguyen}, \citenamefont {Prodhom},
  \citenamefont {Reiher}, \citenamefont {Roux}, \citenamefont {Schlenkrich},
  \citenamefont {Smith}, \citenamefont {Stote}, \citenamefont {Straub},
  \citenamefont {Watanabe}, \citenamefont {Wi\'{o}rkiewicz-Kuczera},
  \citenamefont {Yin},\ and\ \citenamefont {Karplus}}]{CHARMM}%
  \BibitemOpen
  \bibfield  {author} {\bibinfo {author} {\bibfnamefont {A.~D.}\ \bibnamefont
  {MacKerell}}, \bibinfo {author} {\bibnamefont {Jr.}}, \bibinfo {author}
  {\bibfnamefont {D.}~\bibnamefont {Bashford}}, \bibinfo {author}
  {\bibfnamefont {M.}~\bibnamefont {Bellott}}, \bibinfo {author} {\bibfnamefont
  {R.~L.}\ \bibnamefont {Dunbrack}}, \bibinfo {author} {\bibnamefont {Jr.}},
  \bibinfo {author} {\bibfnamefont {J.~D.}\ \bibnamefont {Evanseck}}, \bibinfo
  {author} {\bibfnamefont {M.~J.}\ \bibnamefont {Field}}, \bibinfo {author}
  {\bibfnamefont {S.}~\bibnamefont {Fischer}}, \bibinfo {author} {\bibfnamefont
  {J.}~\bibnamefont {Gao}}, \bibinfo {author} {\bibfnamefont {H.}~\bibnamefont
  {Guo}}, \bibinfo {author} {\bibfnamefont {S.}~\bibnamefont {Ha}}, \bibinfo
  {author} {\bibfnamefont {D.}~\bibnamefont {Joseph-McCarthy}}, \bibinfo
  {author} {\bibfnamefont {L.}~\bibnamefont {Kuchnir}}, \bibinfo {author}
  {\bibfnamefont {K.}~\bibnamefont {Kuczera}}, \bibinfo {author} {\bibfnamefont
  {F.~T.~K.}\ \bibnamefont {Lau}}, \bibinfo {author} {\bibfnamefont
  {C.}~\bibnamefont {Mattos}}, \bibinfo {author} {\bibfnamefont
  {S.}~\bibnamefont {Michnick}}, \bibinfo {author} {\bibfnamefont
  {T.}~\bibnamefont {Ngo}}, \bibinfo {author} {\bibfnamefont {D.~T.}\
  \bibnamefont {Nguyen}}, \bibinfo {author} {\bibfnamefont {B.}~\bibnamefont
  {Prodhom}}, \bibinfo {author} {\bibfnamefont {W.~E.}\ \bibnamefont {Reiher}},
  \bibinfo {author} {\bibfnamefont {B.}~\bibnamefont {Roux}}, \bibinfo {author}
  {\bibfnamefont {M.}~\bibnamefont {Schlenkrich}}, \bibinfo {author}
  {\bibfnamefont {J.~C.}\ \bibnamefont {Smith}}, \bibinfo {author}
  {\bibfnamefont {R.}~\bibnamefont {Stote}}, \bibinfo {author} {\bibfnamefont
  {J.}~\bibnamefont {Straub}}, \bibinfo {author} {\bibfnamefont
  {M.}~\bibnamefont {Watanabe}}, \bibinfo {author} {\bibfnamefont
  {J.}~\bibnamefont {Wi\'{o}rkiewicz-Kuczera}}, \bibinfo {author}
  {\bibfnamefont {D.}~\bibnamefont {Yin}},\ and\ \bibinfo {author}
  {\bibfnamefont {M.}~\bibnamefont {Karplus}},\ }\href@noop {} {\bibfield
  {journal} {\bibinfo  {journal} {J. Phys. Chem. B}\ }\textbf {\bibinfo
  {volume} {102}},\ \bibinfo {pages} {3586} (\bibinfo {year}
  {1998})}\BibitemShut {NoStop}%
\bibitem [{\citenamefont {Yakubovich}\ \emph {et~al.}(2012)\citenamefont
  {Yakubovich}, \citenamefont {Surdutovich},\ and\ \citenamefont
  {Solov'yov}}]{Yakubovich_2012_NIMB.279.135}%
  \BibitemOpen
  \bibfield  {author} {\bibinfo {author} {\bibfnamefont {A.~V.}\ \bibnamefont
  {Yakubovich}}, \bibinfo {author} {\bibfnamefont {E.}~\bibnamefont
  {Surdutovich}},\ and\ \bibinfo {author} {\bibfnamefont {A.~V.}\ \bibnamefont
  {Solov'yov}},\ }\href@noop {} {\bibfield  {journal} {\bibinfo  {journal}
  {Nucl. Instrum. Meth. B}\ }\textbf {\bibinfo {volume} {279}},\ \bibinfo
  {pages} {135} (\bibinfo {year} {2012})}\BibitemShut {NoStop}%
\bibitem [{\citenamefont {de~Vera}\ \emph {et~al.}(2017)\citenamefont
  {de~Vera}, \citenamefont {Surdutovich}, \citenamefont {Mason},\ and\
  \citenamefont {Solov'yov}}]{deVera_2017_EPJD.71.281}%
  \BibitemOpen
  \bibfield  {author} {\bibinfo {author} {\bibfnamefont {P.}~\bibnamefont
  {de~Vera}}, \bibinfo {author} {\bibfnamefont {E.}~\bibnamefont
  {Surdutovich}}, \bibinfo {author} {\bibfnamefont {N.~J.}\ \bibnamefont
  {Mason}},\ and\ \bibinfo {author} {\bibfnamefont {A.~V.}\ \bibnamefont
  {Solov'yov}},\ }\href@noop {} {\bibfield  {journal} {\bibinfo  {journal}
  {Eur. Phys. J. D}\ }\textbf {\bibinfo {volume} {71}},\ \bibinfo {pages} {281}
  (\bibinfo {year} {2017})}\BibitemShut {NoStop}%
\end{thebibliography}%

\end{document}